\documentclass[epjST]{svjour}
\usepackage{graphics}
\usepackage{graphicx}
\usepackage{amsmath}
\usepackage{amssymb}
\usepackage{hyperref}
\usepackage{braket}
\usepackage[percent]{overpic}  

\newcommand{\ba}{\begin{eqnarray}}
\newcommand{\ea}{\end{eqnarray}}
\newcommand{\bsub}{\begin{subequations}}
\newcommand{\esub}{\end{subequations}}

\def\ket#1{|#1\rangle}
\def\bra#1{\langle#1|}
\def\bsu3{\overline{{\rm SU(3)}}}
\def\bso6{\overline{{\rm SO(6)}}}
\def\bPi2{\overline{\Pi}^{(2)}}
\def\blam{\bar{\lambda}}
\def\bmu{\bar{\mu}}
\def\bK{\bar{K}}
\def\b0{\beta_0}
\def\beq{\beta_{\rm eq}}
\def\g0{\gamma_0}
\def\gaeq{\gamma_{\rm eq}}

\begin{document}
\title{SU(3) partial dynamical symmetry and nuclear shapes}
%
\author{A. Leviatan\thanks{\email{ami@phys.huji.ac.il}}}
\institute{Racah Institute of Physics, The Hebrew University,
  Jerusalem 91904, Israel}
\abstract{
We consider several variants of SU(3) partial dynamical
symmetry in relation to quadrupole shapes in nuclei.
Explicit construction of Hamiltonians with such
property is presented in the framework of the
interacting boson model (IBM), including higher order
terms, and in its proton-neutron extension~(IBM-2).
The cases considered include 
a single prolate-deformed shape with solvable ground and
$\gamma$ or $\beta$ bands, coexisting
prolate-oblate shapes with solvable ground bands,
and aligned axially-deformed proton-neutron
shapes with solvable symmetric ground and $\gamma$ bands
and mixed-symmetry scissors and $\gamma$ bands. 
}
\maketitle
\section{Introduction}
\label{sec:intro}

Symmetries play a key role in nuclei by providing
quantum numbers for the classification of states,
determining spectral degeneracies and selection rules, and
facilitating the calculation of matrix elements.
Models based on spectrum generating algebras form
a convenient framework to study their impact and have been used
extensively in nuclear spectroscopy.
Notable examples include
Elliott's SU(3) model~\cite{Elliott58},
symplectic model~\cite{Rowe85},
pseudo SU(3) model~\cite{ps_su3},
monopole and quadrupole pairing models~\cite{GIN},
interacting boson models for
even-even nuclei~\cite{ibm} and boson-fermion models for
odd-mass nuclei~\cite{ibfm}.
In such models, the Hamiltonian is expanded in elements
of a Lie algebra ($G_{\rm dyn}$),
called the dynamical (spectrum generating) algebra,
in terms of which any operator of a physical observable can be
expressed.
A dynamical symmetry (DS) occurs if the Hamiltonian
can be written in terms of the Casimir operators
of a chain of nested algebras~\cite{Iachello15},
$G_{\rm dyn}\!\supset\! G_1 \!\supset\! G_2 \!\supset\!\dots\!
\supset\! G_{\rm sym} $,
terminating in the symmetry algebra $G_{\rm sym}$.
In such a case, the spectrum is completely solvable and
the eigenstates,
$\ket{\lambda_{\rm dyn},\lambda_1,\lambda_2,\ldots,\lambda_{\rm sym}}$,
are labeled by quantum numbers which are the labels of
irreducible representations (irreps) of the algebras
in the chain.
A~given $G_{\rm dyn}$ can encompass several DS chains,
each providing 
characteristic analytic expressions for observables
and definite selection rules for transition processes.
An attractive feature of such models
is that they are amenable to both quantum and classical
treatments.
The classical limit is obtained by introducing coherent
(or intrinsic) states~\cite{Gilmore79,FengGilm90},
which form a basis for studying the geometry of
algebraic models and their relation to intuitive
notions of shapes and excitation modes.

A comprehensive framework for exploring the interplay of
shapes and symmetries in nuclei, is provided by the 
interacting boson model (IBM)~\cite{ibm}.
The latter describes low-lying quadrupole 
collective states in nuclei in terms of $N$ monopole
$(s)$ and quadrupole $(d)$ bosons, representing valence
nucleon pairs. The model is based on a spectrum generating algebra
$G_{\rm dyn}\!=\!{\rm U(6)}$ and a symmetry algebra
$G_{\rm sym}\!=\!{\rm SO(3)}$. The Hamiltonian
is expanded in the elements of U(6),
$\{s^{\dag}s,\,s^{\dag}d_{m},\, d^{\dag}_{m}s,\, d^{\dag}_{m}d_{m '}\}$,
and consists of Hermitian, rotational-scalar interactions 
which conserve the total number of $s$- and $d$- bosons, 
$\hat N \!=\! \hat{n}_s + \hat{n}_d \!=\! 
s^{\dagger}s + \sum_{m}d^{\dagger}_{m}d_{m}$. 
The solvable limits of the IBM correspond to the following
DS chains,
\bsub
\ba
&&\hspace{-0.5cm}
{\rm U(6)\supset U(5)\supset SO(5)\supset SO(3)}
\qquad\;\,\quad\,
\ket{[N], n_d,\tau,n_{\Delta},L} ~,\quad\quad 
\label{U5-ds}
\\
&&\hspace{-0.5cm}
{\rm U(6)\supset SU(3)\supset SO(3)}
\qquad\;\;\qquad\quad\quad\;\,
\ket{[N], (\lambda,\mu),K,L} ~, \quad\quad
\label{SU3-ds}
\\
&&\hspace{-0.5cm}
{\rm U(6)\supset \bsu3\supset SO(3)}
\qquad\;\;\qquad\quad\quad\;\,
  \ket{[N], (\blam,\bmu),\bar{K}, L} ~, \quad\quad
\label{SU3bar-ds}
\\
&&\hspace{-0.5cm}
{\rm U(6)\supset SO(6)\supset SO(5)\supset SO(3)}
\qquad\quad
\ket{[N],\sigma,\tau,n_{\Delta},L} ~. \quad\quad\;
\label{SO6-ds}
\ea
\label{IBMchains}
\esub
Here $N,n_d,(\lambda,\mu),(\blam,\bmu),\sigma,\tau,L$, 
label the relevant irreps of 
U(6), U(5), SU(3), $\bsu3$, SO(6), SO(5), SO(3), 
respectively, and $K,\bar{K},n_{\Delta},$ are multiplicity
labels. The resulting spectra of these DS chains
with leading sub-algebras ${\rm G_1}$: 
U(5), SU(3), ${\rm\overline{SU(3)}}$ and SO(6), 
resemble known paradigms of nuclear collective 
structure:
spherical vibrator, prolate-, oblate- and $\gamma$-soft
deformed rotors, respectively.
Electromagnetic moments and rates can be calculated 
with transition operators of appropriate rank. For example,
the one-body $E2$ operator reads
\ba
\hat{T}(E2) = e_{\rm b} [ \, d^{\dag}s + s^{\dag}\tilde{d}
 + \chi\,(d^{\dag}\tilde{d})^{(2)}\,] ~,
\label{TE2}
\ea
where $\tilde{d}_{m}\!=\! (-1)^{m}d_{-m}$, and standard notation of 
angular momentum coupling is~used.

A geometric visualization of the IBM is obtained by an
energy surface,
\ba
E_{N}(\beta,\gamma) &=&
\bra{\beta,\gamma; N} \hat{H} \ket{\beta,\gamma ; N} ~,
\label{enesurf}
\ea 
defined by the expectation value of the Hamiltonian in
a coherent (intrinsic) state of the form~\cite{gino80,diep80},
\bsub
\ba
\vert\beta,\gamma ; N \rangle &=&
(N!)^{-1/2}[b^{\dagger}_{c}(\beta,\gamma)]^N
\,\vert 0\,\rangle ~,\\
b^{\dagger}_{c}(\beta,\gamma)
&=& (1+\beta^2)^{-1/2}[\beta\cos\gamma 
d^{\dagger}_{0} + \beta\sin{\gamma} 
( d^{\dagger}_{2} + d^{\dagger}_{-2})/\sqrt{2} + s^{\dagger}] ~.
\ea
\label{int-state}
\esub
Here $(\beta,\gamma)$ are
quadrupole shape parameters whose values, $(\beta_{\rm eq},\gamma_{\rm eq})$, 
at the global minimum of $E_{N}(\beta,\gamma)$ define the equilibrium 
shape for a given Hamiltonian. 
The shape can be spherical $(\beta \!=\!0)$ or 
deformed $(\beta \!>\!0)$ with $\gamma \!=\!0$ (prolate), 
$\gamma \!=\!\pi/3$ (oblate), 
$0 \!<\! \gamma \!<\! \pi/3$ (triaxial) or $\gamma$-independent.
The equilibrium deformations associated with the 
DS limits conform with their geometric interpretation 
and are given by,
\bsub
\ba
&&{\rm U(5):} \qquad\;\;\, \beta_{\rm eq}=0 \hspace{2.1cm}
\quad\;\; n_d=0~,\quad\\
&&{\rm SU(3):} \qquad (\beta_{\rm eq} \!=\!\sqrt{2},\gamma_{\rm eq}\!=\!0)
\qquad\;\;\;\, (\lambda,\mu)=(2N,0)~,\quad\\
&&{\bsu3:} \qquad (\beta_{\rm eq} \!=\!\sqrt{2},\gamma_{\rm eq}\!=\!\pi/3)
\qquad (\blam,\bmu)=(0,2N)~,\quad\\
&& {\rm SO(6):} \qquad (\beta_{\rm eq}\!=\!1,\gamma_{\rm eq}\,\,{\rm arbitrary})
\quad\;\; \sigma =N ~.\quad
\ea
\label{int-states-ds}
\esub
The coherent state $\ket{\beq,\gaeq;N}$~(\ref{int-state}),
with the equilibrium deformations, serves as the intrinsic state
for the ground band and is a lowest
(or highest) weight state 
in a particular irrep of the leading sub-algebra 
in each of the chains~(\ref{IBMchains}).
The DS Hamiltonians support a single minimum in their 
energy surface, hence serve as benchmarks for the
dynamics of a single quadrupole shape.

An exact dynamical symmetry (DS) provides considerable
insights into complex dynamics and its merits are
self-evident. 
However, in most applications to realistic systems,
its predictions are rarely fulfilled and one is compelled
to break it.
More often one finds that the assumed symmetry is not obeyed uniformly,
{\it i.e.}, is fulfilled by some of the states but not by others.
The need to address such situations, but still preserve
important symmetry remnants, has motivated
the introduction of partial dynamical symmetry
(PDS)~\cite{lev11}.
The essential idea is to relax the stringent conditions of
{\it complete} solvability so that only part of the eigenspectrum
retains analyticity and/or good quantum numbers.
In the present contribution, we present an explicit
construction of Hamiltonians with the PDS property
in the framework of the IBM,
with applications to nuclear spectroscopy.
We focus the discussion on PDS associated with
SU(3) symmetry which, since the pioneering work by
Elliott~\cite{Elliott58}, has become an essential
ingredient in all shell-model inspired models of
rotational motion in nuclei and
a central theme in nuclear physics~\cite{Kota}.

\section{Partial dynamical symmetry for a single shape}
\label{sec:PDS}

An algorithm for constructing Hamiltonians with a single
PDS has been developed in~\cite{AL92} and further elaborated 
in~\cite{RamLevVan09}. In the IBM, 
the analysis starts from a dynamical symmetry chain,
\ba
   {\rm U(6)\supset G_1\supset G_2\supset \ldots \supset SO(3)}
   &&\;\;\quad
\ket{[N],\lambda_1,\lambda_2,\ldots,L} 
\;\;\quad (\beq,\gaeq) ~,\quad 
\label{u6-ds}
\ea  
with leading sub-algebra $G_1$, related basis 
$\ket{[N],\lambda_1,\lambda_2,\ldots,L}$
and associated shape $(\beta_{\rm eq},\gamma_{\rm eq})$.
A number-conserving Hamiltonian with $G_1$ partial symmetry is found
by writing it as an expansion in terms of $G_1$-tensors
\ba
\hat{H} &=& 
\sum_{\alpha,\beta}u_{\alpha\beta}
\hat{T}^{\dag}_{\alpha}\hat{T}_{\beta} ~,
\label{H-normal}
\ea
which annihilate the states
in a particular $G_1$-irrep, $\lambda_1\!=\!\Lambda_0$,
\ba
\hat{T}_{\alpha}
\ket{[N],\lambda_1\!=\!\Lambda_0,\lambda_2,\ldots,L} = 0 ~.
\label{Talpha1}
\ea
Equivalently, $\hat{T}_{\alpha}$ annihilate the
lowest or highest weight state of this irrep,
\ba
\hat{T}_{\alpha}
\ket{[N],\lambda_1\!=\!\Lambda_0} = 0 ~,
\label{Talpha2}
\ea
from which the different $L$-states
are obtained by projection. If $\lambda_1=\Lambda_0$
is the ground-state irrep, the extremal
state of Eq.~(\ref{Talpha2}) coincides with the
intrinsic state for the ground band,
$\ket{\beq,\gaeq;N}$~(\ref{int-state}),
representing the equilibrium shape $(\beq,\gaeq)$.
The $G_1$-tensors in Eq.~(\ref{H-normal}),
involve $n$-boson creation and annihilation operators
with definite character under the chain~(\ref{u6-ds}).
If both $\hat{T}_{\alpha}$ and the states
$\ket{[N],\lambda_1\!=\!\Lambda_0,\lambda_2,\ldots,L}$
span entire irreps of $G_1$,
then the condition~(\ref{Talpha1})
is satisfied if their Kronecker product does not
contain $G_1$ irreps which belong to the $[N\!-\!n]$ irrep
of U(6).

In general, the Hamiltonian $\hat{H}$ of
Eq.~(\ref{H-normal}) is not invariant under $G_1$,
hence most of its eigenstates
are mixed with respect to $G_1$. However, the
normal-ordered form~(\ref{H-normal}) and the
conditions~(\ref{Talpha1})-(\ref{Talpha2}),
ensure that it has a subset of solvable
zero-energy eigenstates with good
symmetry $G_1$,
implying $G_1$ partial symmetry for $\hat{H}$.
The degeneracy of these states
is lifted, without affecting their wave functions,
by adding to $\hat{H}$ the Hamiltonian,
\ba
\hat{H}_c &=& \sum_{G_i\subset G_1} a_{G_i}\hat{C}[G_i] ~,
\label{H-col}
\ea
composed of the Casimir operators of
sub-algebras of $G_1$ in the chain~(\ref{u6-ds}).
The states $\ket{[N],\lambda_1\!=\!\Lambda_0,\lambda_2,\ldots,L}$
remain solvable eigenstates of the complete Hamiltonian, 
\ba
\hat{H}_{\rm PDS} &=& \hat{H} + \hat{H}_c
= \hat{H}_{\rm DS} + \hat{V}_0  ~,
\label{H-PDS}
\ea
which, by definition, has $G_1$-PDS.
The decomposition of Eq.~(\ref{H-PDS})
is referred to as a resolution of the Hamiltonian into intrinsic
($\hat{H}$) and collective ($\hat{H}_c$) parts~\cite{kirlev85,lev87}.
The former determines the energy surface~(\ref{enesurf})
and band-structure, while the latter
determines the in-band rotational splitting.
For specific choice of parameters, $u_{\alpha\beta}$ in
Eq.~(\ref{H-normal}), $\hat{H}$ reduces to the Casimir
operators of $G_1$, which when combined with $\hat{H}_c$
comprise the DS Hamiltonian ($\hat{H}_{\rm DS}$).
The PDS Hamiltonian ($\hat{H}_{\rm PDS}$) can then be written as
$\hat{H}_{\rm PDS} \!=\!\hat{H}_{\rm DS} \!+\! \hat{V}_0$,
where $V_0$ contains the remaining terms in
$\hat{H}$. In what follows, we review the SU(3)-DS limit
of the IBM, construct Hamiltonians with SU(3)-PDS and
show their relevance to nuclear spectroscopy.

\section{SU(3) dynamical symmetry including higher-order terms}
\label{sec:su3-ds}

The SU(3)-DS limit is appropriate to the dynamics of
a prolate-deformed shape. Its related chain,
quantum numbers and equilibrium deformations are given by
\ba
\hspace{-0.5cm}
{\rm U(6)\supset SU(3)\supset SO(3)} &&\;\;\quad
\ket{[N],(\lambda,\mu),K,L}
\;\;\quad (\beq=\sqrt{2},\gaeq=0) ~.
\label{su3-chain}
\ea
For a given $N$, the allowed SU(3) irreps are 
$(\lambda,\mu)\!=\!(2N \!-\! 4k \!-\! 6m,2k)$ 
with $k,m$, non-negative integers. 
The values of $L$ contained in a given $(\lambda,\mu)$-irrep 
follow the ${\rm SU(3)\supset SO(3)}$ reduction rules~\cite{ibm} 
and the multiplicity label $K$ corresponds geometrically to the
projection of the angular momentum on the symmetry axis.
The states $\ket{[N],(\lambda,\mu),K,L}$ 
form the (non-orthogonal) Elliott basis and the Vergados basis 
$\ket{[N],(\lambda,\mu),\tilde{\chi},L}$
is obtained from it by a standard orthogonalization procedure. 
The two bases coincide in the large-N limit.
The generators of SU(3) are the angular momentum operators,
$L^{(1)}$, and the quadrupole operators, $Q^{(2)}$,
\ba
{\textstyle Q^{(2)}_m = d^{\dag}_{m}s + s^{\dag}\tilde{d}_m 
-\frac{1}{2}\sqrt{7} (d^{\dag}\tilde{d})^{(2)}_m \quad ,\quad
L^{(1)}_m =\sqrt{10}\,(d^{\dagger}\tilde{d})^{(1)}_m} ~.
\label{su3-gen}
\ea
The quadratic and cubic Casimir operators of SU(3)
are given by
\bsub
\ba
\hat C_{2}[{\rm SU(3)}]
&=&
2Q^{(2)}\cdot Q^{(2)}
+ {\textstyle\frac{3}{4}}L^{(1)}\cdot L^{(1)}
\label{C2}\\
\hat C_{3}[{\rm SU(3)}]
&=&
-4\,\sqrt{7}Q^{(2)}\cdot (Q^{(2)}\times Q^{(2)}) ^{(2)}
-{\textstyle\frac{9}{2}\sqrt{3}} Q^{(2)}\cdot 
(L^{(1)}\times L^{(1)})^{(2)} ~,\qquad
\label{C3}
\ea
\label{C2-C3}
\esub
where the centered dot implies a scalar product and
$\hat C_{k}[{\rm G}]$ denotes the Casimir operator of
G of order $k$.
The respective eigenvalues are
\bsub
\ba
f_{2}(\lambda,\mu) &=&
\lambda^2 +\mu^2 +\lambda\mu + 3\lambda + 3\mu ~,
\label{f2}\\
f_{3}(\lambda,\mu) &=&
(\lambda -\mu)(2\lambda+\mu+3)(\lambda+2\mu+3) ~.
\label{f3}
\ea
\label{f2-f3}
\esub
The SU(3) DS Hamiltonian involves a linear combination of
$\hat{C}_{2}[{\rm SU(3)}]$, $\hat{C}_{3}[{\rm SU(3)}]$ and 
$\hat{C}_{2}[SO(3)]=L^{(1)}\cdot L^{(1)}$. The spectrum is
completely solvable and resembles that of an
axially-deformed roto-vibrator composed of
SU(3) $(\lambda,\mu)$-multiplets forming 
rotational bands with $L(L+1)$-splitting. 
The lowest irrep $(2N,0)$ contains the ground band $g(K\!=\!0)$ 
of a prolate deformed nucleus. 
The first excited irrep $(2N\!-\!4,2)$ contains 
both the $\beta(K\!=\!0)$ and $\gamma(K\!=\!2)$ bands. 
States with the same $L$ in different $K$-bands are degenerate. 

Three-body terms allow an additional solvable symmetry-conserving operator, 
\ba
\hat{\Omega} &=& -4\sqrt{3}\, Q^{(2)}\cdot 
(L^{(1)}\times L^{(1)})^{(2)} ~.
\label{Omega}
\ea
This operator is constructed from the SU(3) generators, hence is diagonal 
in $(\lambda,\mu)$. It breaks, however, the aforementioned $K$-degeneracy. 
A well defined procedure exists for obtaining the eigenstates 
of $\hat{\Omega}$ and corresponding 
eigenvalues $\langle \hat{\Omega} \rangle$~\cite{Meyer85,Berghe85}. 
In particular, for the irreps $(\lambda,0)$ and $(\lambda,2)$
with $\lambda$ even, we have
\bsub
\ba
&&
\hspace{-0.5cm}
(\lambda,0)\;  K = 0,\,\; L=0,2,4,\ldots,\lambda:
\qquad\qquad\;\;\;\,
\langle \hat{\Omega}\rangle = (2\lambda +3)L(L+1)~,
\label{ev0}
\\
&&
\hspace{-0.5cm}
(\lambda,2)\;  K= 2,\,\; L=3,5,7,
\ldots,\lambda \!+\! 1,\lambda+2: 
\;\;\;
\langle \hat{\Omega}\rangle = (2\lambda +5)[ L(L+1) - 12]~,\;\;
\label{OmegaLodd}\\
&&
\hspace{-0.5cm}
(\lambda,2)\; K = 0,\,\; L=0: 
\qquad\qquad\qquad\qquad\quad\;\;\;
 \langle \hat{\Omega}\rangle = 0~,\\
&&
\hspace{-0.5cm}
(\lambda,2)\; K = 0,2,\;\;\, L=2,4,6,\ldots,\lambda:
\nonumber\\
&&\;\; \langle \hat{\Omega}\rangle = (2\lambda +5)[ L(L+1) - 6]
\pm 6 \sqrt{(2\lambda+5)^2 + L(L+1)(L-1)(L+2)}~.
\qquad\,
\label{Epm}
\ea
\label{Omegev}
\esub
The two eigenstates $\ket{(\lambda,\mu),L,\pm}$
corresponding to the eigenvalues in Eq.~(\ref{Epm}),
involve a mixture of $K=0,2$, Elliott states
with the $\beta$ band lying above the $\gamma$ band.
Several works have examined the influence of the
operator $\hat{\Omega}$ (\ref{Omega})
on nuclear spectra,  within the
IBM~\cite{Berghe85,Bona88,Vant90} and the symplectic 
shell model~\cite{RDW84,DR85}.

\section{SU(3) partial dynamical symmetry for a single shape}
\label{sec:su3-PDS}

According to Section~\ref{sec:PDS},
the method to construct Hamiltonians with SU(3)-PDS
is based on the identification of SU(3) tensors
which annihilate states in a given 
SU(3) irrep $(\lambda,\mu)$, 
chosen here to be the ground band irrep $(2N,0)$. 
The tensors involve $n$-boson
operators with definite character under the SU(3)
chain~(\ref{su3-chain}),
\ba
\hat B^\dag_{[n](\lambda,\mu)\tilde{\chi}\ell m},\;\;
\tilde B_{[n^5](\mu,\lambda)\tilde{\chi}\ell m}\equiv
(-)^{m}
(\hat B^\dag_{[n](\lambda,\mu)\tilde{\chi}\ell,-m})^{\dag} ~.
\;\;
\ea
The SU(3) tensor operators for $n$=2, span the irrep
$(\lambda,\mu)=(0,2)$ and are given by,
\ba
B^{\dag}_{[2](0,2)0;00} &=& \tfrac{1}{3\sqrt{2}}\,P^{\dag}_{0}
\;\; , \;\;
B^{\dag}_{[2](0,2)0;2m} = \tfrac{1}{3\sqrt{2}}\,P^{\dag}_{2m} ~.
\label{B2}
\ea
They are constructed of boson-pair operators with
angular momentum $\ell =0,\,2$,
\bsub
\ba
P^{\dagger}_{0} &=& d^{\dag}\cdot d^{\dag} - 2(s^{\dagger})^2 ~,
\label{P0}\\
P^{\dagger}_{2m} &=& 2d^{\dag}_{m}s^{\dag} + 
\sqrt{7}\, (d^{\dag} d^{\dag})^{(2)}_{m} ~.
\label{P2}
\ea
\label{PL}
\esub
The SU(3) tensor operators for $n$=3, span the irreps
$(\lambda,\mu)=(2,2)$ and $(0,0)$,
\ba
\hat B^\dag_{[3](2,2)0;00} &=& \tfrac{1}{30}W^{\dag}_{0}
\;\; , \;\;\,
\hat B^\dag_{[3](2,2)2;2m} = \tfrac{1}{\sqrt{78}}W^{\dag}_{2m}
\;\; , \;\;
\hat B^\dag_{[3](2,2)0;2m} = \tfrac{1}{\sqrt{910}}V^{\dag}_{2m} ~,
\quad
\nonumber\\
\hat B^\dag_{[3](2,2)2;3m} &=& \tfrac{1}{\sqrt{30}}W^{\dag}_{3m}
\;\; , \;\;
\hat B^\dag_{[3](2,2)2;4m} = \tfrac{1}{\sqrt{30}}W^{\dag}_{4m}
\;\; , \;\;
\hat B^\dag_{[3](0,0)0;00} = \tfrac{1}{18}\Lambda^{\dag} ~,\quad
\label{B3}
\ea
and are obtained by coupling $s^{\dag}$ and $d^{\dag}_m$
to the $n=2$ SU(3) tensors of Eq.~(\ref{PL}),
\ba
&&W^{\dag}_{0} = 5P^{\dag}_{0}s^{\dag} - P^{\dag}_{2}\cdot d^{\dag} 
\;\; , \;\;
W^{\dag}_{2m} = P^{\dag}_{0}d^{\dag}_{m} + 2P^{\dag}_{2m}s^{\dag} 
\;\; , \;\;
V^{\dag}_{2m} = 6P^{\dag}_{0}d^{\dag}_{m} - P^{\dag}_{2m}s^{\dag} ~,
\qquad
\nonumber\\
&&W^{\dag}_{3m} = (P^{\dag}_{2}d^{\dag})^{(3)}_{m}
\;\; , \;\;
W^{\dag}_{4m} = (P^{\dag}_{2}d^{\dag})^{(4)}_{m} 
\;\; , \;\;
\Lambda^{\dag} =
P^{\dag}_{0}s^{\dag} \!+\! P^{\dag}_{2}\cdot d^{\dag} ~.\qquad
\label{WV}
\ea
It should be noted that
$P^{\dag}_{0}d^{\dag}_{m} + P^{\dag}_{2m}s^{\dag} \!=\!
-\tfrac{\sqrt{7}}{2} (P^{\dag}_{2}d^{\dag})^{(2)}_{m}$
and $W^{\dag}_{3m} \!=\!
{\textstyle\sqrt{7}}(d^{\dag}d^{\dag})^{(2)}d^{\dag})^{(3)}_m$.

The Hermitian conjugate of the operators
in Eq.~(\ref{PL}), $P_0$ and $P_{2m}$, transform
as $(2,0)$ under SU(3) and satisfy
\bsub
\ba
P_{0}\,\ket{[N],(2N,0),K=0, L} &=& 0 ~,
\label{P0vanish}\\
P_{2m}\,\ket{[N],(2N,0),K=0, L} &=& 0 ~, 
\;\;\;\;\;
L=0,2,4,\ldots, 2N~.
\label{P2vanish}
\ea
\label{P0P2vanish}
\esub
These relations follow from the fact that the action
of the operators $P_{\ell m}$ leads to a state with 
$N\!-\!2$ bosons in the U(6) irrep $[N\!-\!2]$, 
which does not contain the SU(3) irreps obtained
from the product $(2,0)\times (2N,0)$.
The indicated $L$-states in Eq.~(\ref{P0P2vanish}) span the
entire SU(3) irrep $(\lambda,\mu)=(2N,0)$
and form the rotational members of the ground band.
They can be obtained by SO(3) projection from
the intrinsic state for the SU(3) ground band,
\ba
\ket{g;(2N,0)K=0;N} = \ket{\beq=\sqrt{2},\gaeq=0 ; N} ~,
\label{int-g}
\ea
defined as in Eq.~(\ref{int-state})
with the SU(3) equilibrium deformations
and representing a prolate-deformed shape.

The relations of Eq.~(\ref{P0P2vanish}) ensure that
the Hermitian conjugate of all SU(3) tensors in
Eqs.~(\ref{B2}) and (\ref{B3}),
$P_0,\,P_{2m},\,\,W_{\ell m},\,V_{2m},\,\Lambda$,
annihilate $\ket{g;(2N,0)K\!=\!0;N}$ and the corresponding
$L$-projected states. By the procedure of
Eq.~(\ref{H-normal}), the most general (2+3)-body
Hamiltonian with SU(3) partial symmetry
can be written as,
\ba
\hat H &=&
h_2\,P^\dag_2\!\cdot\! \tilde P_2+h_0\,P^\dag_0 P_0+
t_0^a\,\Lambda^\dag\Lambda+
t_0^b\,W_0^\dag W_0+
t_0^c\,(\Lambda^\dag W_0+W_0^\dag\Lambda)\qquad\quad
\nonumber\\&&+
t_2^a\,V_2^\dag\!\cdot\!\tilde V_2+
t_2^b\,W_2^\dag\!\cdot\!\tilde W_2+
t_2^c\,(V_2^\dag\!\cdot\!\tilde W_2+W_2^\dag\!\cdot\!\tilde V_2)
+t_3\,W_3^\dag\!\cdot\!\tilde W_3
+t_4\,W_4^\dag \cdot\tilde W_4~,\qquad
\label{e_pds}
\ea
and has a subset of SU(3) basis states,
$\ket{[N],(2N,0),K\!=\!0,L}$, as zero-energy eigenstates.
Several SU(3)-preserving interactions
are contained in the expression~(\ref{e_pds}). 
Specifically, the (integrity basis) term,
\ba
\hat{\Omega} - (4\hat{N}+3)L^{(1)}\cdot L^{(1)} 
&=& 
2(\hat{N}-2)P^{\dag}_{0}P_{0}
-2(2\hat{N}-1)P^{\dag}_{2}\cdot\tilde{P}_{2} 
+ 2 \Lambda^{\dag}\Lambda
\nonumber\\
&&\;
-2( \Lambda^{\dag}sP_{0} +
P^{\dag}_{0}s^{\dag}\Lambda )
+2\,W^{\dag}_{2}\cdot \tilde{W}_{2} 
+4\,W^{\dag}_{3}\cdot \tilde{W}_{3} ~,\qquad
\label{Omegap}
\ea
and the following SU(3)-scalar terms,
\bsub
\ba
&&\hat{\theta}_2 \equiv
-\hat C_{2}[{\rm SU(3)}] + 2\hat N (2\hat N+3) =
P^{\dagger}_{0}P_{0} + P^{\dagger}_{2}\cdot \tilde{P}_{2} ~,
\label{theta2}\\
&&(\hat{N}-2)\hat{\theta}_2 = 
6 [\,5\sum_{[\tilde{\chi};L]}
B^{\dag}_{[3](2,2)\tilde{\chi};L}\cdot \tilde{B}_{[3^5](2,2)\tilde{\chi};L}
+9B^{\dag}_{[3](0,0)0;0}\tilde{B}_{[3^5](0,0)0;0}\, ] ~,
\qquad\quad\\
&& \hat{C}_{3}[{\rm SU(3)}] +
(2\hat{N}+3)\left [ 3\hat{\theta}_2 - 2\hat N (4\hat{N}+3)\right ]
= 2\Lambda^{\dag}\Lambda ~. \qquad
\label{LamLam}
\ea 
\label{su3-scalar}
\esub
The three terms in Eq.~(\ref{su3-scalar})
are related to the Casimir operators of SU(3),
Eq.~(\ref{C2-C3}), hence are naturally included in the
dynamical symmetry Hamiltonian. Allowing $N$-dependent
coefficients, the latter can be transcribed in the form,
\bsub
\ba
\hat{H}_{\rm DS} &=& 
h_1\,\Lambda^{\dag}\Lambda +
h_2\,\hat{\theta}_2 + 
C\,\hat{C}_{2}[\rm SO(3)] ~.
\label{H-DS}\\
E_{\rm DS} &=& [h_2 +\tfrac{3}{2}h_1\,(2N+3)]
\,[-f_2(\lambda,\mu)+ 2N(2N+3)\,]
\nonumber\\
&&
+\tfrac{1}{2}h_1\, [\,f_3(\lambda,\mu) - 2N(2N+3)(4N+3)\,]
+ C\,L(L+1) ~.\quad
\label{E-DS}
\ea
\esub
The SU(3)-PDS Hamiltonian can then be written
as in Eq.~(\ref{H-PDS}),
\ba
\hat{H}_{\rm PDS} &=& 
\hat{H}_{\rm DS} + h_3\, P^{\dagger}_{0}P_{0} + 
h_{4} P^{\dag}_{0}s^{\dag}sP_{0}
+h_{5}\left ( \Lambda^{\dag}sP_{0} + P^{\dag}_{0}s^{\dag}\Lambda\right )
\nonumber\\
&&
+h_{6}\,W^{\dag}_{2}\cdot \tilde{W}_{2} 
+h_{7}\,W^{\dag}_{3}\cdot \tilde{W}_{3} 
+h_{8}\,s^{\dag}P^{\dag}_{2}\cdot\tilde{P}_{2}s ~.
\label{HPDShi}
\ea

In general, $\hat{H}_{\rm PDS}$~(\ref{HPDShi}) does not
preserve SU(3) yet, by construction, 
for {\em any} choice of parameters, it has a solvable
ground band with good SU(3) symmetry,
\bsub
\ba
g:
&&\ket{[N],(2N,0),K=0,L} \;\;\;\; L=0,2,4,\ldots, 2N\\ 
&& E_{\rm PDS}=CL(L+1) ~.
\ea
\label{EgL}
\esub
For specific choices, additional solvable states are obtained.
One class of SU(3)-PDS Hamiltonians is obtained by choosing in
Eq.~(\ref{HPDShi}) the $h_3,h_4,h_5$ terms, involving the operators
$P^{\dag}_0$~(\ref{PL}) and $\Lambda^{\dag}$~(\ref{WV}).
$P_0$ satisfies Eq.~(\ref{P0vanish}) and in addition,
\ba
P_{0}\,\ket{[N],(\lambda,\mu)\!=\!(2N-4k,2k),K\!=\!2k,L}
= 0 ~,\;\; L=K,K+1,\ldots, (2N-2k)~.\quad
\label{P0k}
\ea
For $k> 0$, the indicated $L$-states
span only part of the SU(3) irreps 
$(\lambda,\mu)=(2N-4k,2k)$ and form the rotational members
of excited $\gamma^{k}(K\!=\!2k)$ bands.
This result follows from the fact that $P_{0}$ 
annihilates the intrinsic states of these bands, 
\ba
\ket{\gamma^k_{K=2k}; (2N-4k,2k)K\!=\!2k);N} &=&
\sqrt{\tfrac{3^k(2N-4k+1)!!}{k!(2N-2k+1)!!}}
(B^{\dag}_{\gamma, 2})^{k}\,\ket{g; N-2k} ~,
\label{int-gam}
\ea
where $B^{\dag}_{\gamma,2} = \tfrac{1}{3\sqrt{2}}P^{\dag}_{2,2}$
and $\ket{g; N-2k}$ is obtained from Eq.~(\ref{int-g}).
The operator $\Lambda$ transforms as $(0,0)$ under SU(3) and
annihilates {\it all} states in the irreps $(2N-4k,2k)$,
\ba
\Lambda\,\ket{[N],(2N-4k,2k),K,L} &=& 0 ~.
\label{Lam0}
\ea
This result follows from the fact that the U(6) irrep $[N-3]$ 
does not contain SU(3) irreps obtained from the product 
$(0,0)\times (2N-4k,2k)$.
The relations in Eqs.~(\ref{P0k}) and (\ref{Lam0}) ensure that the
following Hamiltonian,
\ba
\hat{H}_{\rm PDS} &=& \hat{H}_{DS} 
+h_{3}\,P^{\dag}_{0}P_{0} + 
h_{4}\, P^{\dag}_{0}s^{\dag}sP_{0}
+ h_{5}\,\left ( \Lambda^{\dag}sP_{0} + 
P^{\dag}_{0}s^{\dag}\Lambda\right ) ~,
\label{hPDS-1}
\ea
has solvable ground band $g(K\!=\!0)$, Eq.~(\ref{EgL}), and
$\gamma^{k}(K\!=\!2k)$ bands with good SU(3) symmetry,
\bsub
\ba
(\gamma^{k})_{K\!=\!2k}: &&
\ket{[N],(2N-4k,2k),K\!=\!2k,L}
\;\;
L\!=\!K,K\!+\!1,K\!+\!2,\ldots, (2N\!-\!2k),\qquad \\
&& E_{PDS} = 
h_{2}\,6k(2N -2k+1) + CL(L+1) ~.
\ea
\label{gamma-PDS}
\esub
The remaining eigenstates of $\hat{H}_{PDS}$, 
in particular, members of the $\beta(K\!=\!0)$ band, are mixed. 
\begin{figure}[t]
\begin{minipage}{16.5pc}
\fbox{\includegraphics[width=\linewidth,clip=]{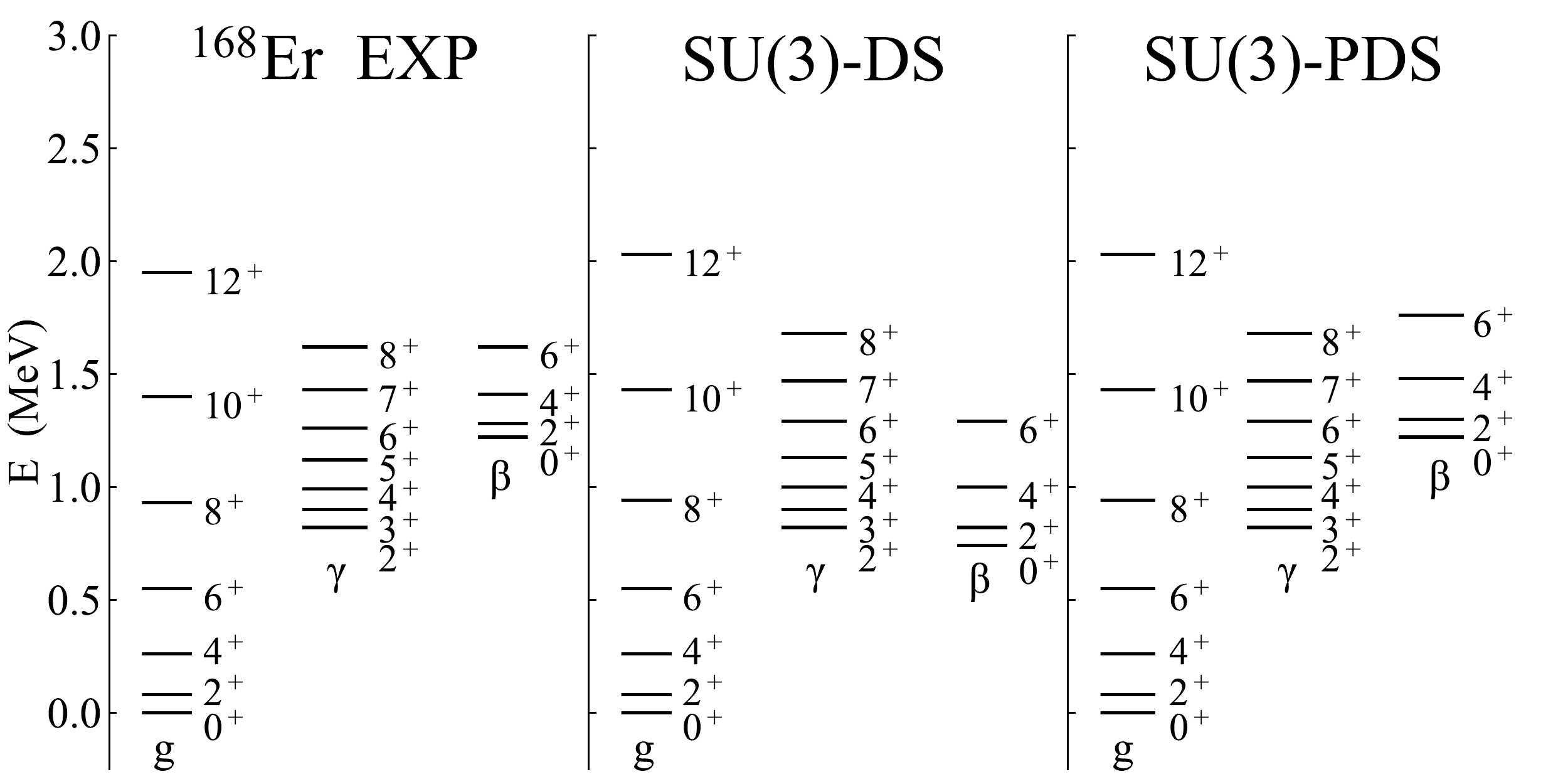}}
\caption{\small\label{fig1-Er168spec}
Observed spectrum of $^{168}$Er compared with SU(3)-DS and
SU(3)-PDS calculations. The latter employ
$\hat{H}_{\rm DS}$~(\ref{H-DS}) and
$\hat{H}_{\rm PDS}$~(\ref{hPDS-1})
with $h_2\!=\!4,\,C\!=\!13,\,h_3\!=\!4$ keV
(other $h_i\!=\!0$)
and $N\!=\!16$. Adapted from~\cite{Lev96}.}
\end{minipage}\hspace{0.6cm}%
\hspace{-0.2cm}%
\begin{minipage}{14pc}
\includegraphics[width=\linewidth,clip=]{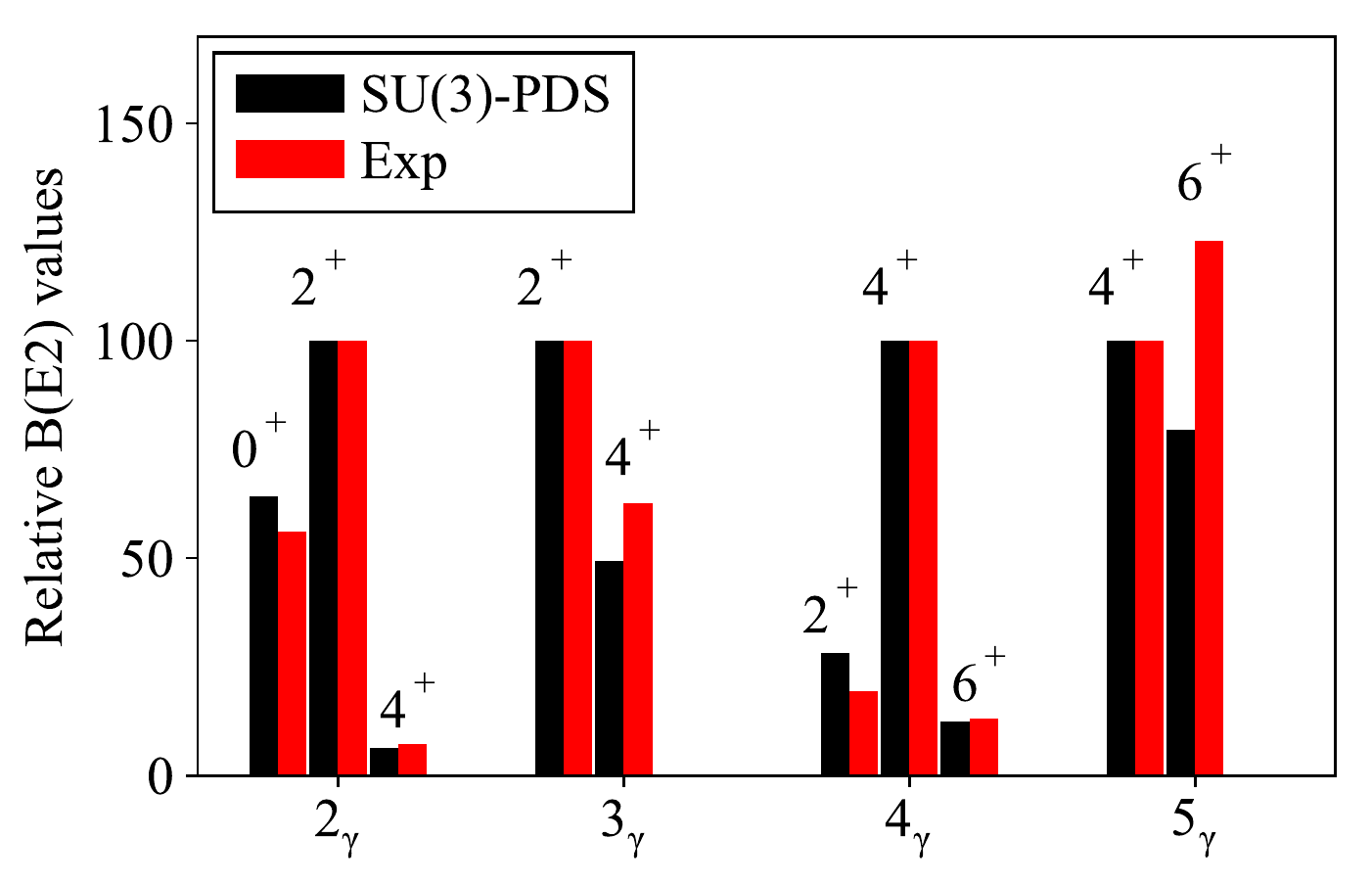}
\caption{\small\label{fig2-Er168be2}
Comparison of the SU(3)-PDS parameter-free predictions with
    the data on the
relative $B(E2; L_{\gamma}\to L)$ values for $\gamma\to g$ $E2$ transitions
in $^{168}$Er. Adapted from ~\cite{Casten14}.}
\label{figGd156-new}
\end{minipage}
\end{figure}

The experimental spectra
of the  ground $g(K\!=\!0)$, $\gamma(K\!=\!2)$ and $\beta(K\!=\!0)$ bands
in $^{168}$Er
is shown in Fig.~\ref{fig1-Er168spec}, and compared with an
exact DS and PDS calculations~\cite{Lev96}, employing the
Hamiltonian~(\ref{hPDS-1}).
The SU(3) PDS spectrum is clearly seen to be an
improvement over the
exact SU(3) DS description, since the $\beta$-$\gamma$
degeneracy of the latter is lifted.
The ground and gamma are still pure SU(3) bands, but
the beta
band is found to contain $13\%$ admixtures into the
dominant $(2N-4,2)$ irrep~\cite{LevSin99}.
Since the wave functions of the solvable states are known,
one has at hand
{\it analytic} expressions for matrix 
elements of observables between them~\cite{Isacker83}.
The $E2$ operator~(\ref{TE2}), can be transcribed as
$\hat{T}(E2) \!=\!
\alpha Q^{(2)} + \theta (d^{\dag}s + s^{\dag}\tilde{d})$,
with $Q^{(2)}$ defined in Eq.~(\ref{su3-gen}).
Since the solvable ground and gamma bands reside in
different SU(3) irreps, 
$Q^{(2)}$ as a generator, cannot connect them and,
consequently, $B(E2)$ ratios for $\gamma\to g$
transitions do no depend on the $E2$
parameters ($\alpha,\theta$) nor on parameters of the
PDS Hamiltonian~(\ref{hPDS-1}).
Overall, as shown in Fig.~\ref{fig2-Er168be2},
these parameter-free predictions of SU(3)-PDS
account well for the data in
$^{168}$Er~\cite{Lev96,Casten14}.
Similar evidence for this type of SU(3)-PDS has
been presented for other rare-earth and actinide
nuclei~\cite{Casten14,Couture15,Casten16},
suggesting a wider applicability of this concept.

Another class of Hamiltonians with SU(3) PDS is obtained by
choosing the $h_6,h_7$ terms in Eq.~(\ref{HPDShi}),
involving the operators
$W^{\dag}_{2 m}$ and $W^{\dag}_{3 m}$~(\ref{WV}).
$W_{\ell m}$ ($\ell \!=\! 2,3$) annihilate the
states of the ground band, $\ket{[N],(2N,0),K=0,L}$,
and in addition satisfy,
\ba
W_{\ell m}\,
\ket{[N],(\lambda,\mu)\!=\!(2N-4,2),K\!=\!0,L} = 0
\;\;, \;\;
L=0,2,4,\ldots, (2N-4) ~.
\label{WL}
\ea
The indicated $L$-states span part of the irrep
$(2N-4,2)$ of SU(3) and comprise the $\beta(K\!=\!0)$ band.
This result follows from the fact that
$W_{\ell m}$ annihilate the intrinsic state of this band, 
\ba
\ket{\beta; (2N-4,2)K=0; N} = 
\sqrt{\tfrac{3}{2N - 1}}\,
B^{\dag}_{\beta}\ket{g; N-2}
\;\; ,\;\; B^{\dag}_{\beta}
= \tfrac{1}{3\sqrt{6}}(\sqrt{2}P^{\dag}_{0} - P^{\dag}_{2,0}) ~,
\quad
\label{int-bet}
\ea
where $\ket{g; N-2}$ is obtained from Eq.~(\ref{int-g}).
This property ensures that the following
Hamiltonian~\cite{levramisa13}, 
\ba
\hat{H}_{\rm PDS} =
\hat{H}_{\rm DS}+
h_6\,W_2^\dag\!\cdot\!\tilde{W}_2
+ h_7\,W_3^\dag\!\cdot\!\tilde{W}_3 ~,
\label{hPDS-2}
\ea
has a solvable ground band $g(K=0)$, Eq.~(\ref{EgL}),
and a solvable $\beta(K=0)$ band,
\bsub
\ba
\beta: &&
\ket{[N],(2N-4,2),K\!=\!0,L}
\;\;\;\; L=0,2,4,\ldots, 2N\\
&& E_{PDS} = 
h_{2}\,6(2N -1) + CL(L+1) ~.
\ea
\label{beta-PDS}
\esub
Other bands, in particular the $\gamma$ band, are mixed.
For the special case, $h_7 = 2h_6$, $\hat{H}_{\rm PDS}$~(\ref{hPDS-2})
has additional solvable states, 
$\ket{[N],(2N-4,2),K\!=\!2,L}$ with $L\!=\!3,5,7,\ldots, (2N-3),2N-2$, 
and energy $h_2\,6(2N -1) + h_6\, 3 [\,8(N-1)^2 - L(L+1)]$,
a result obtained by combining
Eqs.~(\ref{OmegaLodd}) and (\ref{Omegap}).

\begin{figure}[t]
\begin{minipage}{16.5pc}
\includegraphics[width=\linewidth,clip=]{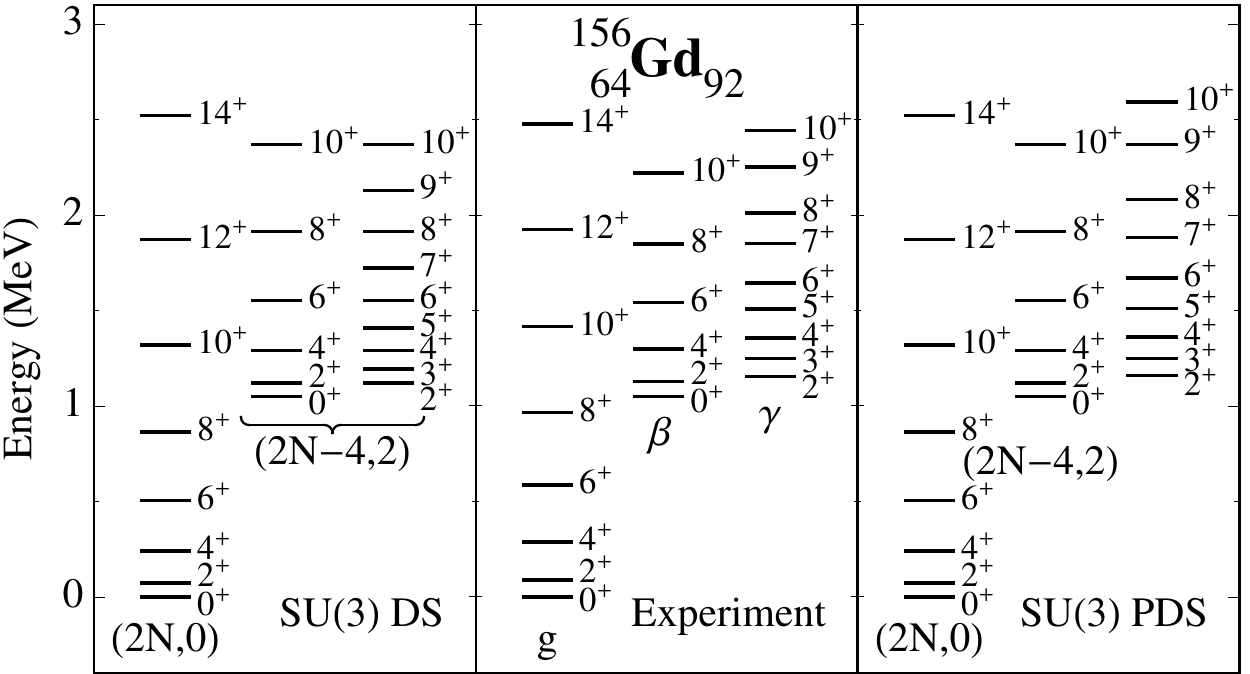}
\caption{
\small\label{fig3-Gd156spec}
Observed spectrum of $^{156}$Gd
compared with SU(3)-DS and SU(3)-PDS calculations.
The latter employ
$\hat{H}_{\rm DS}$~(\ref{H-DS}) and
$\hat{H}_{\rm PDS}$~(\ref{hPDS-2})
with $h_1\!=\!0,\,h_2\!=\! 7.6,\,C\!=\! 12.0,\,
h_6\!=\! -0.23,\, h_7\!=\! 1.54$ keV and $N=12$.
Adapted from~\cite{levramisa13}.}
\end{minipage}\hspace{0.4cm}%
\begin{minipage}{14pc}
\includegraphics[width=\linewidth,height=3cm,clip=]{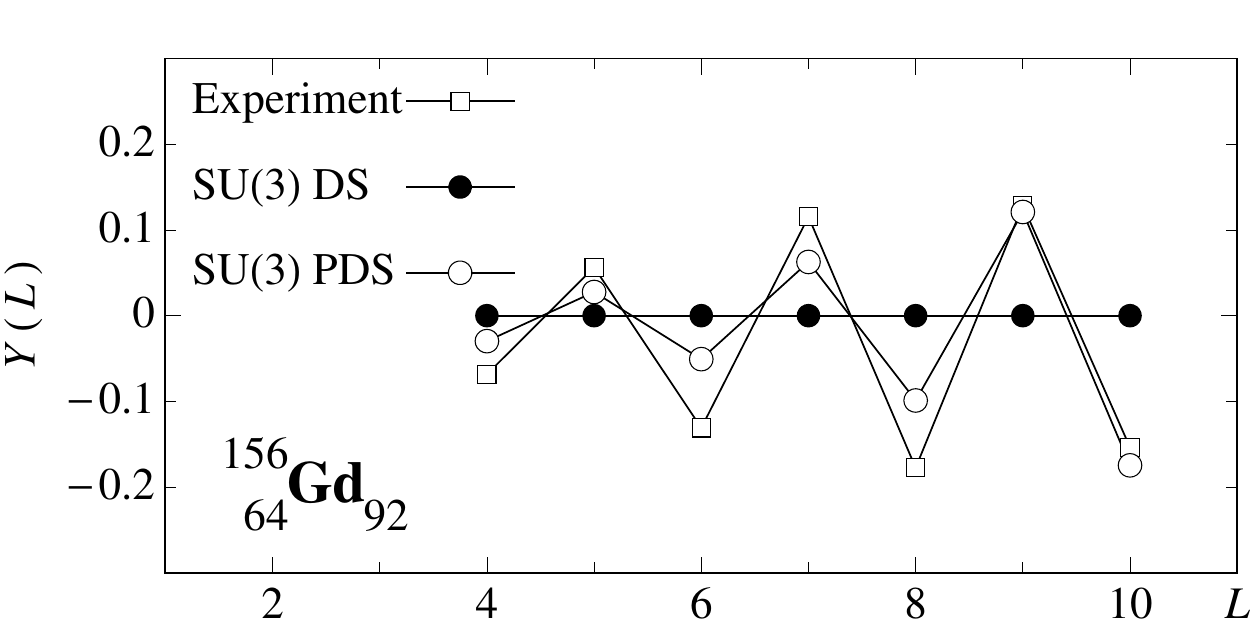}
\caption{
\small\label{fig4-Gd156gam-stag}
Observed and calculated (SU(3)-DS and SU(3)-PDS)
odd-even staggering of the $\gamma$ band in $^{156}$Gd.
Here
$Y(L)\!=\!\frac{2L-1}{L}\times\frac{E(L)-E(L-1)}{E(L)-E(L-2)}-1$,
where $E(L)$ is the energy
of a $\gamma$-band level with angular momentum $L$.
For SU(3)-DS, $Y(L)\!=\!0$, as expected of a rigid rotor.
Adapted 
from~\cite{levramisa13}.}
\end{minipage}
\end{figure}

A comparison, in Fig.~\ref{fig3-Gd156spec},
of the experimental spectrum of $^{156}$Gd
with an SU(3)-DS calculation,
shows a good description for properties of 
states in the ground and $\beta$ bands,
however, the resulting fit to energies of the $\gamma$-band
is quite poor. The latter are not degenerate with the $\beta$ band and,
moreover, display an odd-even staggering with pronounced deviations 
from a rigid-rotor $L(L+1)$ pattern.
This effect can be visualized by plotting the quantity
$Y(L)$, defined in the caption of Fig.~\ref{fig4-Gd156gam-stag}.
For a rotor this quantity is flat, $Y(L)=0$,
as illustrated with the SU(3) DS calculation, 
which is in marked disagreement with the empirical data. 
In the PDS calculation, the gamma band contains
$15\%$ SU(3) admixtures into the dominant $(2N-4,2)$ irrep 
and the empirical odd-even staggering is well reproduced~\cite{levramisa13}.
The PDS results for the $\gamma$ band are obtained without 
affecting the solvability and SU(3) purity of states 
in the ground and beta bands. Since the wave functions of the
latter states are known, one has at hand analytic expressions for
$E2$ transitions between them~\cite{Isacker83},
that can be used to test the SU(3)-PDS.
As shown in Table~\ref{tab-be2},
recent measurements of lifetimes for $E2$ decays from states of
the $\beta$-band in $^{156}$Gd~\cite{Aprahamian18},
confirm the PDS predictions.
\begin{table*}
\centering
\caption{\small\label{tab-be2}
Observed and calculated $B(E2)$ values in $^{156}$Gd.
For both the SU(3) DS and PDS calculations,
the parameters of the E2 operator, Eq.~(\ref{TE2}), are
$e_{\rm b}=0.166$~$eb$ and $\chi=-0.168$~\cite{levramisa13}.
The indicated ranges in the
experimental $B(E2)$ values for
transitions from the $\beta$ band, reflect uncertainties
in the lifetime measurements~\cite{Aprahamian18}.
}
\begin{tabular}{llll|lllll}
\noalign{\smallskip}\hline\noalign{\smallskip}
$L^{\pi}_i$ & $L^{\pi}_f$ & Experiment & PDS\hspace {0.5cm} & 
$L^{\pi}_i$ & $L^{\pi}_f$ & Experiment & PDS\\
\hline
$  2^+_1$  & $0^+_1$ & 0.933~{\sl 25}  & 0.933 &
$0^+_\beta$ & $2^+_1$ & 0.021 $\rightarrow$ 0.055  & 0.034 & \\ 
$  4^+_1$  & $2^+_1$ & 1.312~{\sl 25}  & 1.313 &
$2^+_\beta$ & $0^+_1$ & 0.0028 $\rightarrow$ 0.0057  & 0.0055 & \\ 
$  6^+_1$  & $4^+_1$ & 1.472~{\sl 40}  & 1.405 &
$2^+_\beta$ & $2^+_1$& 0.016 $\rightarrow$ 0.031 & 0.0084 & \\ 
$  8^+_1$  & $6^+_1$& 1.596~{\sl 85}  & 1.409 &
$2^+_\beta$ & $4^+_1$& 0.018 $\rightarrow$ 0.037 & 0.020 &\\ 
$10^+_1$   & $8^+_1$& 1.566~{\sl 70}  & 1.364 &
$4^+_\beta$ & $2^+_1$& 0.0047 $\rightarrow$ 0.011 &  0.0067 \\
 & & & &
$4^+_\beta$ & $4^+_1$& 0.0098 $\rightarrow$ 0.022 & 0.0067 \\
 & & & &
$4^+_\beta$ & $6^+_1$& 0.0080 $\rightarrow$ 0.018 & 0.021 \\
 & & & &
$4^+_\beta$ & $2^+_\beta$& 1.00 $\rightarrow$ 2.19 &  0.951 & \\ 
\noalign{\smallskip}\hline
\end{tabular}
\end{table*}

\section{Multiple partial dynamical symmetries and shape coexistence}
\label{sec:mult-pds}

Coexistence of different shapes in the same nucleus
is a ubiquitous phenomena, known to occur widely 
across the nuclear chart~\cite{Heyde11}.
It entails the presence in the spectrum of several states
(or bands of states) at similar energies with distinct properties,
reflecting the nature of their dissimilar dynamics.
In the shell model description of nuclei near
shell-closure, this phenomena is attributed to the occurrence
of multi-particle multi-hole intruder excitations across
shell gaps. In a mean-field based approach,
the coexisting shapes are associated with different minima 
of an energy surface calculated self-consistently.
The relevant Hamiltonians contain competing terms
with different tendencies,
resulting in a shape-phase transition in which the two minima
cross and the underlying configurations interchange
their roles.

\begin{figure}[t!]
\centering
\hspace{-0.7cm}
\includegraphics[width=6cm]{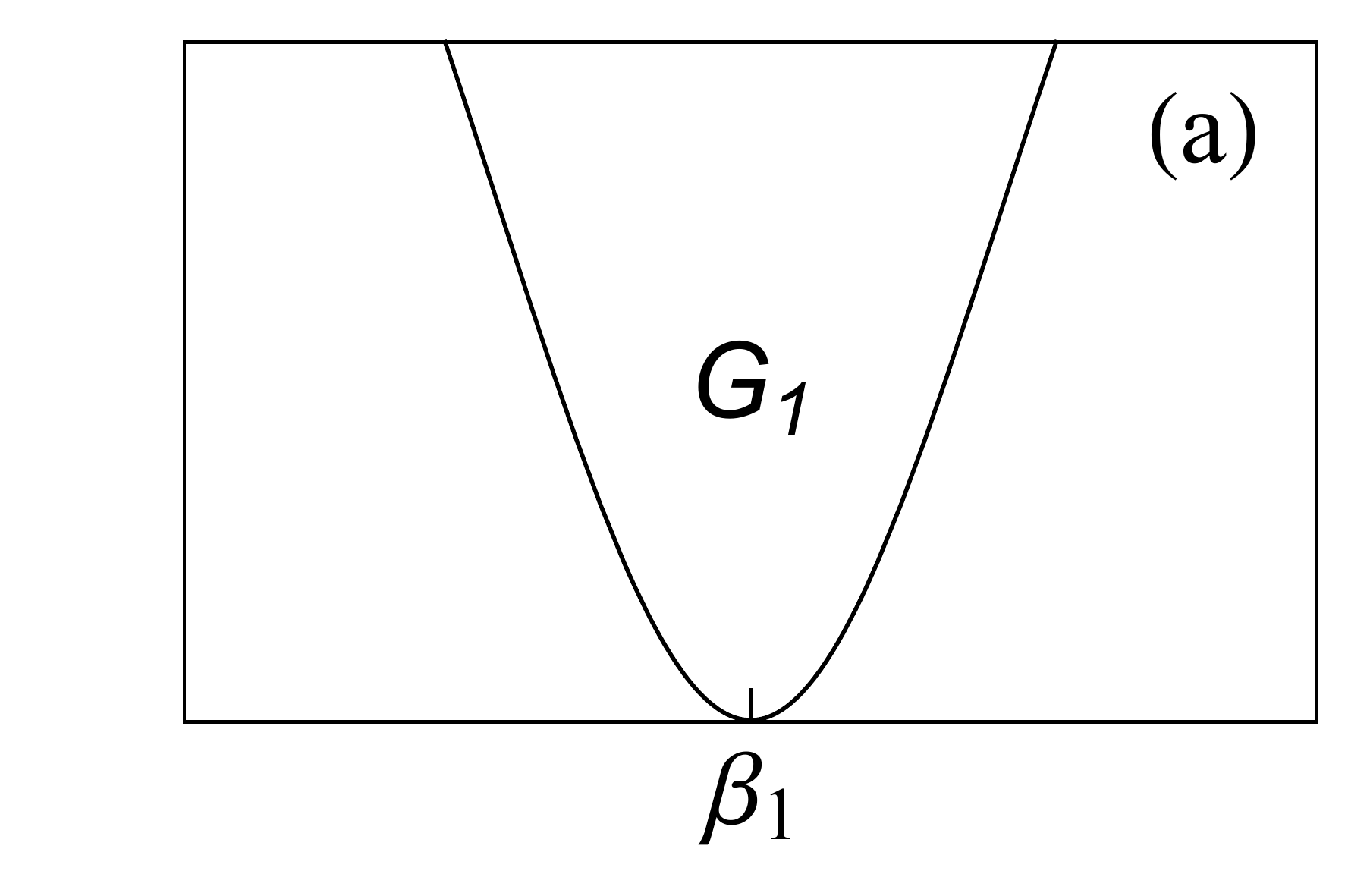}
\hspace{-0.4cm}
\includegraphics[width=6cm]{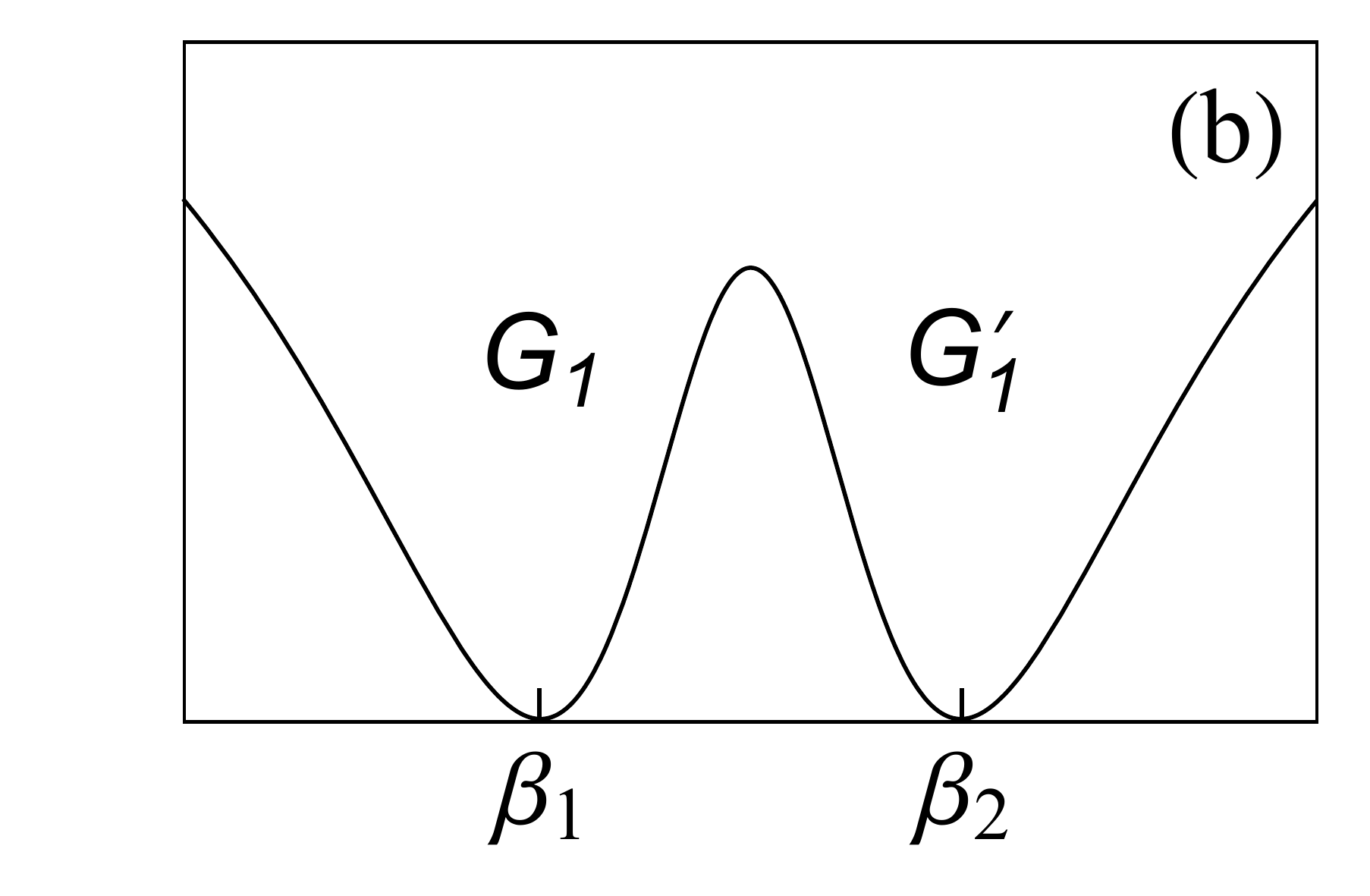}
\caption{
\small\label{fig5-Emult-min}
Energy surfaces accommodating 
(a)~a single minimum, associated with single $G_1$-DS
or $G_1$-PDS; 
(b)~double minima, associated with coexisting $G_1$-PDS and $G_1'$-PDS.}
\end{figure}

From an algebraic perspective, the dynamics of a single shape
is associated with a single dynamical symmetry,
either exact or partial.
The corresponding DS or PDS Hamiltonians support
a single minimum in their energy energy surface at
particular deformations.
In a similar spirit, multiple shapes are associated
with different dynamical symmetries
(denoted by $G_1,$ and $G_{1}'$ in Fig.~\ref{fig5-Emult-min}).
The relevant Hamiltonians
support multiple minima in their energy surface at
particular deformations and mix
these incompatible (non-commuting) symmetries.
In such circumstances, exact DSs are broken, and any
remaining symmetries can at most be partial, {\it i.e.},
shared by only a subset of states.
To explore such scenarios of persisting symmetries in
relation to shape coexistence, 
requires an extension of the PDS algorithm
of Section~\ref{sec:PDS},
to encompass a construction of Hamiltonians with
several distinct PDSs~\cite{Leviatan07,LevDek16,LevGav17}.
We focus on the dynamics in the vicinity of the critical 
point, where the corresponding multiple minima in the energy surface 
are near-degenerate and the structure changes most rapidly.

For that purpose, consider two different shapes 
specified by equilibrium deformations 
($\beta_1,\gamma_1$) and ($\beta_2,\gamma_2$) 
whose dynamics is described, respectively, by the
following DS chains
\bsub
\ba
{\rm U(6)\supset G_1\supset G_2\supset \ldots \supset SO(3)} &&\;\;\quad
\ket{[N],\, \lambda_1,\,\lambda_2,\,\ldots,\,L} 
\;\;\qquad (\beta_1,\gamma_1) ~,\quad 
\label{ds-G1}\\
{\rm U(6)\supset G'_1\supset G'_2\supset \ldots \supset SO(3)} &&\;\;\quad
\ket{[N],\, \sigma_1,\,\sigma_2,\,\ldots,\,L} 
\;\;\qquad (\beta_2,\gamma_2) ~,\quad 
\label{ds-G1prime}
\ea
\label{ds-G1G1p}
\esub
with different leading sub-algebras ($G_1\neq G'_1$) and
associated bases. As portrayed in Fig.~\ref{fig5-Emult-min},
at the critical point, the corresponding minima representing the
two shapes, and respective ground bands are degenerate. 
Accordingly, we consider an intrinsic critical-point Hamiltonian
as in Eq.~(\ref{H-normal}), $\hat{H} = \sum_{\alpha,\beta}u_{\alpha\beta}
\hat{T}^{\dag}_{\alpha}\hat{T}_{\beta}$,
but now require the operators $\hat{T}_{\alpha}$ to satisfy 
simultaneously the following two conditions,
\bsub
\ba
\hat{T}_{\alpha}\ket{[N],\lambda_1\!=\Lambda_0,\lambda_2,\ldots,L} 
&=& 0 ~,
\label{basis1}\\
\hat{T}_{\alpha}\ket{[N],\sigma_1=\Sigma_0,\sigma_2,\ldots,L} 
&=&0 ~.
\label{basis2}
\ea
\label{bases12}
\esub
The states of Eq.~(\ref{basis1}) reside in the $\lambda_1\!=\!\Lambda_0$ irrep 
of $G_1$, are classified according to the DS-chain (\ref{ds-G1}), hence 
have good $G_1$ symmetry. Similarly,  
the states of Eq.~(\ref{basis2}) reside in the $\sigma_1\!=\!\Sigma_0$ irrep 
of $G'_1$, are classified according to the DS-chain (\ref{ds-G1prime}), 
hence have good $G'_1$ symmetry.
Equivalently, $\hat{T}_{\alpha}$ annihilate the
extremal states of both irreps,
\bsub
\ba
\hat{T}_{\alpha}
\ket{[N],\lambda_1\!=\!\Lambda_0} = 0 ~,
\label{lw1}\\
\hat{T}_{\alpha}
\ket{[N],\sigma_1\!=\!\Sigma_0} = 0 ~,
\label{lw2}
\ea
\label{lw12}
\esub
from which the different $L$-states
are obtained by projection.
Although $G_1$ and $G'_1$ are incompatible, 
both sets of states in Eqs.~(\ref{bases12}) are eigenstates of the
same Hamiltonian. In general, $\hat{H}$ itself is not necessarily 
invariant under $G_1$ nor under $G'_1$ and, therefore, its other
eigenstates can be mixed with respect to both $G_1$ and $G'_1$.
If $\lambda_1\!=\!\Lambda_0$ and
$\sigma_1\!=\!\Sigma_0$
are the ground-state irreps of $G_1$ and $G'_1$,
the extremal states of Eqs.~(\ref{lw12}) coincide with the
intrinsic states $\ket{\beta_1,\gamma_1;N}$
and $\ket{\beta_2,\gamma_2;N}$ , Eq.~(\ref{int-state}),
of the two ground bands, representing the equilibrium shapes
$(\beta_1,\gamma_1)$ and $(\beta_2,\gamma_2)$, respectively.
Identifying, as in Eq.~(\ref{H-col}),
the collective part of the Hamiltonian ($\hat{H}_c$),
with the Casimir operators of the algebras which are
common to both chains~(\ref{ds-G1G1p}), in particular
$\hat{C}_{2}[\rm SO(3)]$,
the two sets of states in Eqs.~(\ref{bases12})
remain (non-degenerate)
eigenstates of the complete critical-point Hamiltonian
which has the form as in Eq.~(\ref{H-PDS}),
$\hat{H}_{\rm PDS} \!=\! \hat{H} + \hat{H}_c
\!=\! \hat{H}_{\rm DS} + \hat{V}_0$.
The latter, by construction, has both $G_1$-PDS and $G'_1$-PDS.
In what follows, we apply the above procedure to a case study of
coexisting axially-deformed shapes with multiple PDSs.

\section{SU(3) partial dynamical symmetry and prolate-oblate coexistence}
\label{su3-po}

The DS limits appropriate to prolate and oblate shapes correspond,
respectively, to the chains~\cite{ibm},
\bsub
\ba
{\rm U(6)\supset SU(3)\supset SO(3)} &&\;\;\quad
\ket{[N],(\lambda,\mu),K, L}  
\;\;\qquad
(\beta_{\rm eq}=\sqrt{2},\gamma_{\rm eq}=0) ~,\quad
\label{SU3}
\\
{\rm U(6)\supset \bsu3\supset SO(3)} &&\;\;\quad
\ket{[N],(\blam,\bmu),\bar{K}, L}
\;\;\qquad
(\beta_{\rm eq}=\sqrt{2},\gamma_{\rm eq}=\pi/3) ~.\qquad
\label{SU3bar}
\ea
\label{su3chains}
\esub
The two chains have similar properties, but the
classification of states is different.
For a given $N$, the allowed
$\bsu3$ irreps are 
$(\blam,\bmu)\!=\!(2k,2N\!-\!4k\!-\!6m)$,
with $k,m$, non-negative integers, 
compared to $(\lambda,\mu)\!=\!(2N \!-\! 4k \!-\! 6m,2k)$
for SU(3). The $\bsu3$ algebra involves the following quadrupole
and angular momentum operators,
\ba
{\textstyle \bar{Q}^{(2)}_m = d^{\dag}_{m}s + s^{\dag}\tilde{d}_m 
+\frac{1}{2}\sqrt{7} (d^{\dagger}\tilde{d})^{(2)}_m \quad ,\quad
L^{(1)} =\sqrt{10}(d^{\dagger}\tilde{d})^{(1)}_m} ~.
\label{bsu3-gen}
\ea
The generators of SU(3) and $\bsu3$, $Q^{(2)}$~(\ref{su3-gen}) 
and $\bar{Q}^{(2)}$~(\ref{bsu3-gen}), 
and corresponding basis states,
$\ket{[N],(\lambda,\mu),K,L}$ and $\ket{[N],(\blam,\bmu),\bar{K},L}$, 
are related by a change of phase
$(s^{\dag},s)\rightarrow (-s^{\dag},-s)$, 
induced by the operator 
\ba
{\cal R}_s=\exp(i\pi\hat{n}_s) \;\; , \;\;
\hat{n}_s=s^{\dag}s ~.
\label{Rs}
\ea
The quadratic and cubic Casimir operators of $\bsu3$,
$\hat C_{2}[{\rm \bsu3}]$ and $\hat C_{3}[{\rm \bsu3}]$,
have the form as in Eqs.~(\ref{C2-C3})
with  $Q^{(2)}$ replaced by $\bar{Q}^{(2)}$ and eigenvalues
$f_{2}(\blam,\bmu)$ and $f_{3}(\blam,\bmu)$, Eqs.~(\ref{f2-f3}).
As previously mentioned, in the SU(3)-DS, 
the prolate ground band  $g(K\!=\!0)$ has 
$(2N,0)$ character and the $\beta(K\!=\!0)$ and $\gamma(K\!=\!2)$ 
bands have $(2N\!-\!4,2)$. 
In the $\bsu3$-DS, the oblate ground band $g(\bK\!=\!0)$ has 
$(0,2N)$ character and the excited $\beta(\bK\!=\!0)$ and 
$\gamma(\bK\!=\!2)$ bands
have $(2,2N\!-\!4)$. 
Henceforth, we denote such prolate and oblate bands by 
$(g_1,\beta_1,\gamma_1)$ and ($g_2,\beta_2,\gamma_2$), respectively. 
Since ${\cal R}_sQ^{(2)}{\cal R}_s^{-1} \!=\! -\bar{Q}^{(2)}$, 
the SU(3) and $\bsu3$ DS spectra are identical but
the quadrupole moments of corresponding states differ. 

\begin{figure}[t]
\centering
\includegraphics[width=\linewidth]{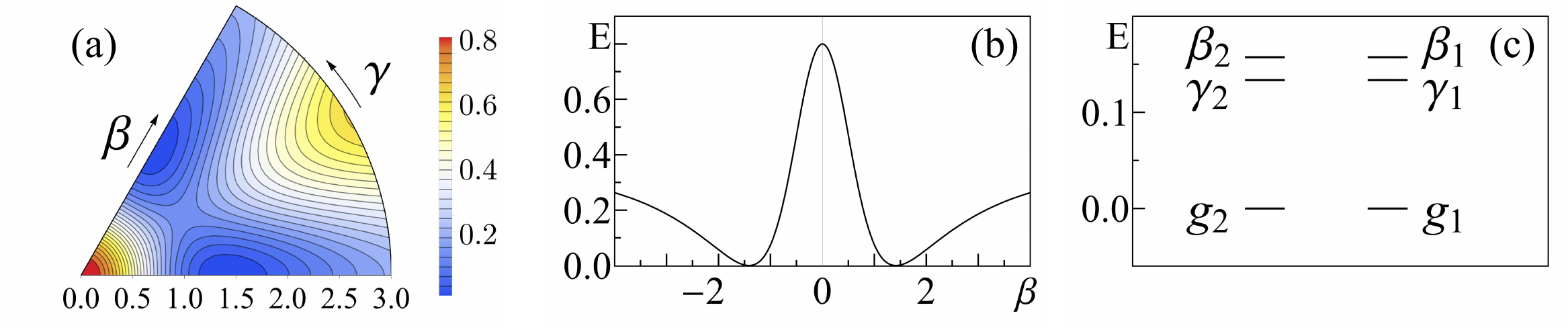}
\caption{\label{fig6-PO}
\small
Prolate-oblate shape coexistence.
(a)~Contour plots of the energy surface~(\ref{surfacePO}),   
(b)~$\gamma\!=\!0$ sections, and
(c)~bandhead spectrum, for the Hamiltonian
$\hat{H}'$ of Eq.~(\ref{HprimePO}),
with parameters
$r_0\!=\!0.2,\,r_2\!=\!0.4,\,r_3\!=\!0.567,\,C\!=\!0,\,
\alpha\!=\!0.018$ and $N\!=\!20$,
with multiple SU(3)-PDS and $\bsu3$-PDS.
Adapted from~\cite{LevGav17}.} 
\end{figure}

Following the procedure of Section~\ref{sec:mult-pds},
the intrinsic part of the critical-point Hamiltonian, 
relevant to prolate-oblate coexistence, is constructed
of operators which annihilate both the SU(3) and
$\bsu3$ ground bands. Specifically,
$P^{\dag}_0$~(\ref{P0}), which is a $(0,2)$ [$(2,0)$]
tensor under SU(3) [$\bsu3$] and $W^{\dag}_{3m}$~(\ref{WV}),
which is a $(2,2)$ tensor under both algebras. These tensors
commute with ${\cal R}_s$~(\ref{Rs}) and satisfy,
\bsub
\ba
&& P_0\,\ket{[N],(\lambda,\mu)\!=\!(2N,0),K\!=\!0,L}
= W_{3m}\,\ket{[N],(\lambda,\mu)\!=\!(2N,0),K\!=\!0,L}
= 0 ~,\qquad\;\;\;
\label{p0-w3-su3}\\
&& P_0\,\ket{[N],(\blam,\bmu)\!=\!(0,2N),\bar{K}\!=\!0,L}
= W_{3m}\,\ket{[N],(\blam,\bmu)\!=\!(0,2N),\bar{K}\!=\!0,L}
= 0 ~.\qquad\;\;\;
\label{p0-w3-bsu3}
\ea
\label{p0-w3} 
\esub
with $L=0,2,4,\ldots 2N$.
The states of Eq.~(\ref{p0-w3-su3})
comprise the SU(3) prolate ground band ($g_1$), projected
from the intrinsic state,
$\ket{\beq\!=\!\sqrt{2},\gaeq\!=\!0;N}$, Eq.~(\ref{int-state}).
Similarly, the states of Eq.~(\ref{p0-w3-bsu3})
comprise the $\bsu3$ oblate ground band ($g_2$), projected from
the intrinsic state $\ket{\beq\!=\!\sqrt{2},\gaeq\!=\!\pi/3;N}$.
The latter deformations are equivalent to
$\textstyle{(\beq\!=\!-\sqrt{2},\gaeq\!=\!0)}$, since the
intrinsic states $\ket{\beta,\gamma\!=\!\pi/3;N}$ and
$\ket{-\beta,\gamma\!=\!0;N}$
are related by a rotation by Euler angles
$\textstyle{(\pi/2,\pi/2,\pi/2)}$, hence contain the same $L$-states.
The resulting intrinsic Hamiltonian is found to be~\cite{LevDek16},
\ba
\hat{H} = 
r_0\,P^{\dag}_0\hat{n}_sP_0 + r_2\,P^{\dag}_0\hat{n}_dP_0 
+r_3\,W^{\dag}_3\cdot\tilde{W}_3 ~.
\label{HintPO}
\ea
The corresponding energy surface, 
$E_{N}(\beta,\gamma) = N(N-1)(N-2)\tilde{E}(\beta,\gamma)$, 
is given by
\ba
\tilde{E}(\beta,\gamma) = 
\left\{(\beta^2-2)^2
\left [r_0 + r_2\beta^2\right ] 
+r_3 \beta^6(1-\Gamma^2)\right \}(1+\beta^2)^{-3} ~.
\label{surfacePO}
\ea
The surface is an even function of $\beta$ and 
$\Gamma \!=\! \cos 3\gamma$. For $r_0,r_2,r_3\geq 0$, 
$\hat{H}$ is positive definite and 
$\tilde{E}(\beta,\gamma)$ has two degenerate global minima, 
$(\beta\!=\!\sqrt{2},\gamma\!=\!0)$ and 
$(\beta\!=\!\sqrt{2},\gamma\!=\!\pi/3)$ 
[or equivalently $(\beta\!=\!-\sqrt{2},\gamma\!=\!0)$], at $\tilde{E}=0$.
$\beta\!=\!0$ is always an extremum, which is a local minimum (maximum) for 
$r_2 \!>\! 4r_0$ ($r_2 \!<\! 4r_0$), at $\tilde{E}=4r_0$.
Additional extremal points include saddle points at 
$[\beta_1\!>\!0,\gamma\!=\!0,\pi/3]$, $[\beta_2\!>\!0,\gamma\!=\!\pi/6]$ 
and a local maximum at $[\beta_3\!>\!0,\gamma\!=\!\pi/6]$. 
The saddle points, when exist, support
a barrier separating the various minima, as seen in Fig.~\ref{fig6-PO}. 
For large $N$, the normal modes involve 
$\beta$ and $\gamma$ vibrations about the 
respective deformed minima, with frequencies
\bsub
\ba
&&
\epsilon_{\beta 1}=\epsilon_{\beta 2} 
= \frac{8}{3}(r_0+ 2r_2)N^2 ~,
\\
&&\epsilon_{\gamma 1}=\epsilon_{\gamma 2} = 4r_3N^2 ~.
\ea
\label{d-modes}
\esub

Identifying the collective part with $\hat{C}_2[{\rm SO(3)}]$, 
we arrive at the following complete Hamiltonian, 
\ba
\hat{H}_{\rm PDS} &=& 
r_0\,P^{\dag}_0\hat{n}_sP_0 + r_2\,P^{\dag}_0\hat{n}_dP_0 
+r_3\,W^{\dag}_3\cdot\tilde{W}_3
+ C\,\hat{C}_2[\rm SO(3)] ~.\qquad
\label{HPDS-po}
\ea
$\hat{H}_{\rm PDS}$ is not invariant under SU(3) nor $\bsu3$,
yet the relations of Eqs.~(\ref{p0-w3}) ensure that it has
a solvable prolate ground band $g_1(K=0)$ with good SU(3) symmetry
and simultaneously, an oblate ground band, $g_2(\bar{K}=0)$ with
good $\bsu3$ symmetry,
\bsub
\ba
g_1: && \ket{[N],(\lambda,\mu)\!=\!(2N,0),K=0,L}
\;\;\;\; L=0,2,4,\ldots, 2N ~,
\label{Eg1L}\\
g_2: && \ket{[N],(\blam,\bmu)\!=\!(0,2N),\bar{K}=0,L} ~,
\label{Eg2L}\\
&& E_{g1}(L) = E_{g2}(L) = C\,L(L+1) ~.
\label{Eg1LEg2L}
\ea
\label{g1g2}
\esub
$\hat{H}_{\rm PDS}$~(\ref{HPDS-po}) thus exhibits both
SU(3)-PDS and $\bsu3$-PDS.
It has, however, an undesired feature that it
commutes with the ${\cal R}_{s}$ operator
of Eq.~(\ref{Rs}). Consequently, all non-degenerate eigenstates
have well-defined $s$-parity. 
This implies vanishing quadrupole moments for an $E2$ 
operator which is odd under a change of sign of the $s$-bosons,
{\it e.g.}, for $\chi=0$ in Eq.~(\ref{TE2}).
To overcome this difficulty, we add a small $s$-parity breaking
term and consider the Hamiltonian,
\ba
\hat{H}' &=& \hat{H}_{\rm PDS} + \alpha\,\hat{\theta}_2 ~,\qquad
\label{HprimePO}
\ea
where ${\textstyle\hat{\theta}_2}$ is defined
in Eq.~(\ref{theta2}). The added term
contributes to $\tilde{E}(\beta,\gamma)$~(\ref{surfacePO})
a~small component 
${\textstyle\tilde{\alpha}(1+\beta^2)^{-2}[ 
(\beta^2\!-\!2)^2 \!+\! 2\beta^2(2 \!-\!2\sqrt{2}\beta\Gamma 
\!+\!\beta^2)]}$, 
with ${\textstyle\tilde{\alpha}=\alpha/(N-2)}$. 
The linear $\Gamma$-dependence distinguishes 
the two deformed minima and slightly lifts
their degeneracy, as well as that of the normal modes~(\ref{d-modes}). 

\begin{figure}[t]
\begin{minipage}{20pc}
\vspace{0.5cm}
\includegraphics[width=\linewidth,clip=]{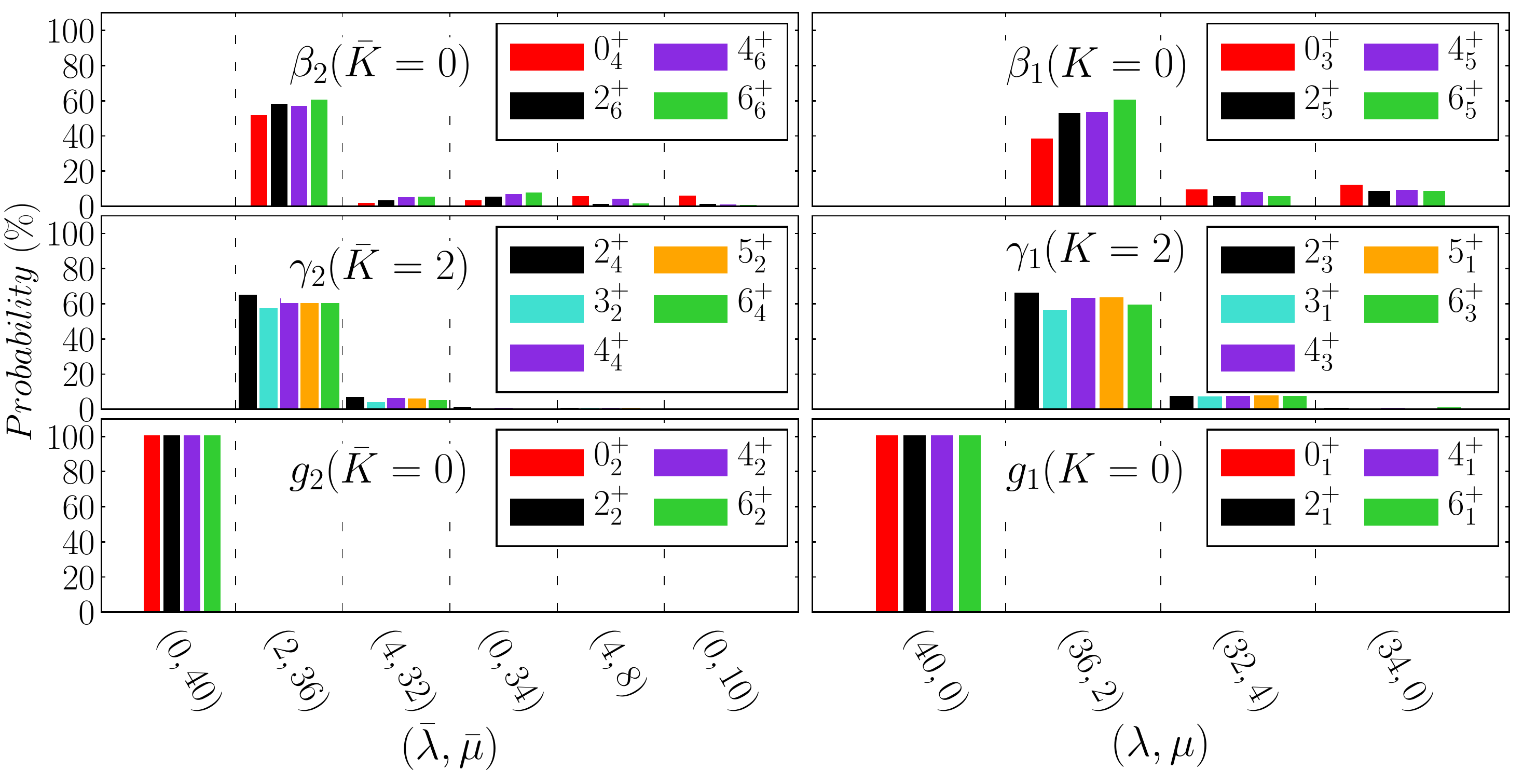}
\hspace{-0.3cm}%
\caption{\small\label{fig7-decomp}
SU(3) $(\lambda,\mu)$- and $\bsu3$ $(\blam,\bmu)$-decompositions 
for members of the prolate ($g_1,\beta_1,\gamma_1$) 
and oblate ($g_2,\beta_2,\gamma_2$) bands, eigenstates of 
$\hat{H}'$ (\ref{HprimePO}) with parameters as in Fig.~\ref{fig6-PO}, 
resulting in prolate-oblate shape coexistence. 
Shown are probabilities larger than 5\%.
Note the purity of the prolate ground band $g_1(K=0)$
[oblate ground band $g_2(\bar{K}=0)$]
with respect to SU(3) [$\bsu3$].
Adapted from~\cite{LevGav17}.}
\end{minipage}\hspace{0.6cm}%
\hspace{-0.29cm}%
\begin{minipage}{10pc}
\centering
\includegraphics[width=0.8\linewidth,clip=]{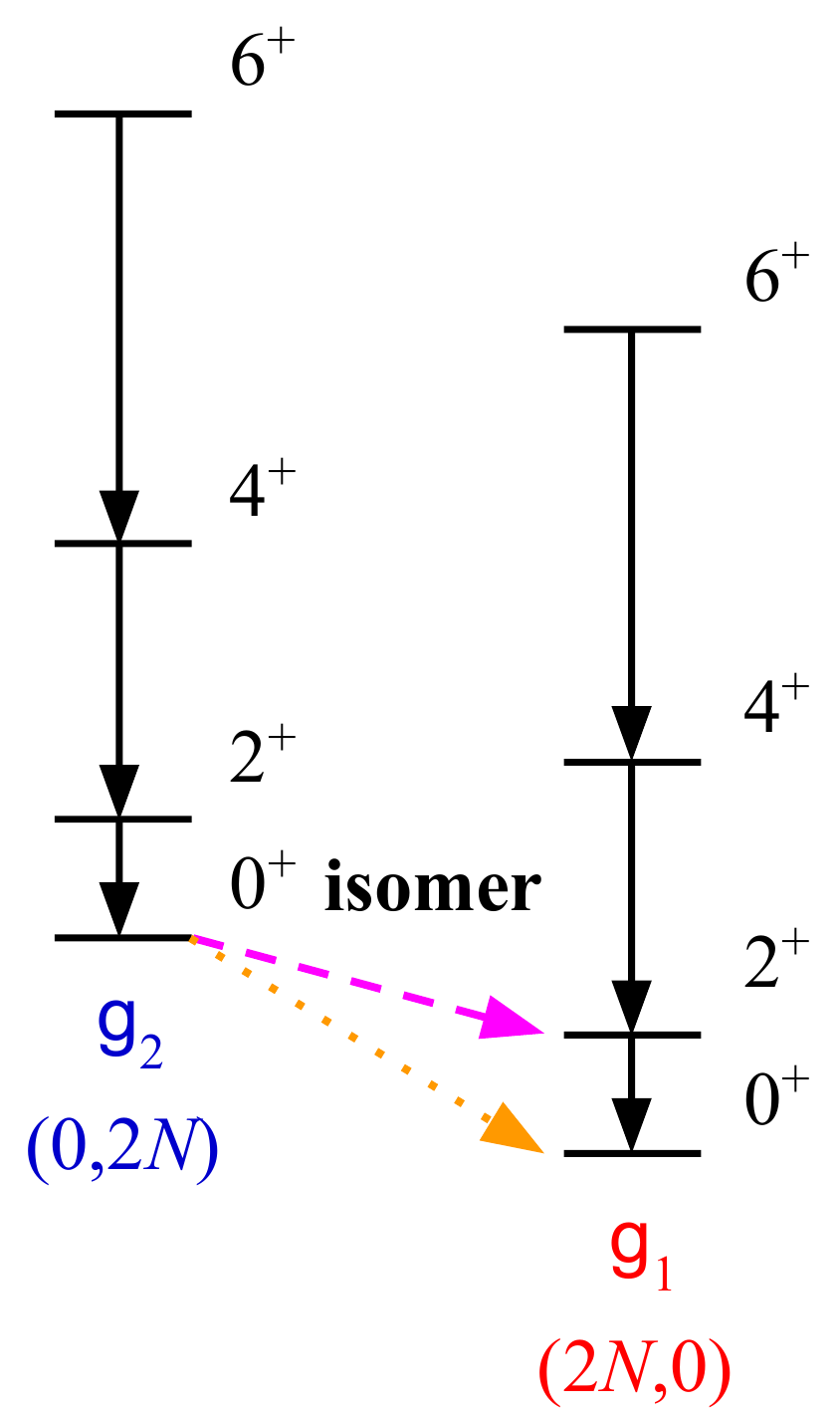}
\caption{\small\label{fig8-isomer}
SU(3) and $\bsu3$ PDSs. Strong intraband $E2$ transitions
(solid lines) obey Eq.~(\ref{be2}).
Retarded $E2$ and $E0$ decays (dashed and dotted lines)
identify isomeric states.}
\end{minipage}
\end{figure}

Fig.~\ref{fig6-PO} shows $\tilde{E}(\beta,\gamma)$,
$\tilde{E}(\beta,\gamma\!=\!0)$ 
and the bandhead spectrum of $\hat{H}'$ (\ref{HprimePO}). 
The prolate $g_1$-band 
remains solvable with energy $E_{g1}(L)$, Eq.~(\ref{Eg1LEg2L}).
The oblate $g_2$-band experiences a slight shift of
order ${\textstyle\tfrac{32}{9}\alpha N^2}$ and
displays a rigid-rotor like spectrum. 
Replacing ${\textstyle\hat{\theta}_2}$ 
by ${\textstyle\bar{\theta}_2 \!=\! 
  -\hat{C}_{2}[\bsu3] + 2\hat{N}(2\hat{N}+3)}$,
reverses the sign of the linear $\Gamma$ term in the energy surface 
and leads to similar effects, but 
interchanges the role of prolate and oblate bands
in the spectrum of $\hat{H}'$.
The SU(3) and $\bsu3$ decompositions in Fig.~\ref{fig7-decomp}
demonstrate that these bands are pure DS basis states, with 
$(2N,0)$ and $(0,2N)$ character, respectively, while excited
$\beta$ and $\gamma$ bands exhibit considerable mixing.
The critical-point Hamiltonian thus has a subset of states with
good SU(3) symmetry, a~subset of states with good $\bsu3$ symmetry
and all other states are mixed with respect to both SU(3) and
$\bsu3$.
These are precisely the defining ingredients of SU(3)-PDS
coexisting with $\bsu3$-PDS.

Since the wave functions for the 
members of the $g_1$ and $g_2$ bands 
are known, one can derive closed expressions for their 
quadrupole moments and $E2$ rates,
which are the observables most closely related to the nuclear shapes.
For the general $E2$ operator of Eq.~(\ref{TE2}),
the quadrupole moments are found to be
\ba
Q_L &=& 
{\textstyle
\mp e_{\rm b}\,\sqrt{\frac{16\pi}{40}}\frac{L}{2L+3}
\left [\frac{4(2N-L)(2N+L+1)}{3(2N-1)}
(1 \pm \bar{\chi})
  \mp\bar{\chi}\, (4N+3)\right ]} ~,
\label{quadmom}
\ea
where ${\textstyle\bar{\chi}=\frac{2}{\sqrt{7}}\chi}$,
and the upper (lower) signs correspond to the
prolate-$g_1$ (oblate-$g_2$) band. The $B(E2)$ values for
transitions within these bands, are found to be
\ba
&&B(E2; g_i,\, L+2\to g_i,\,L) = 
\nonumber\\
&&
\quad
\;\;
{\textstyle
e_{\rm b}^2\,\frac{3(L+1)(L+2)}{2(2L+3)(2L+5)}
\frac{(2N-L)(2N+L+3)}{2}
\left [1 -\frac{2(N-1)}{3(2N-1)}
(1\pm\bar{\chi})\right ]^2} ~,
\qquad\qquad
\label{be2}
\ea
where the upper (lower) sign corresponds to
$g_1\to g_1$ ($g_2\to g_2$) transitions.
It should be noted that in
Eqs.~(\ref{quadmom})-(\ref{be2}), the factor 
${\textstyle(1 \pm \bar{\chi})}$ vanishes for
$\textstyle{\chi= \mp\frac{1}{2}\sqrt{7}}$
for which the $E2$ operator coincides with the
SU(3) or $\bsu3$ generator $Q^{(2)}$~(\ref{su3-gen}) or
$\bar{Q}^{(2)}$~(\ref{bsu3-gen}).

The purity and good quantum numbers of selected states 
enable the derivation of symmetry-based selection rules for 
electromagnetic transitions (notably, for $E2$ and $E0$ decays) and the 
subsequent identification of isomeric states.
The general $E2$ operator $\hat{T}(E2)$ of Eq.~(\ref{TE2}),
can be written as a sum of $Q^{(2)}$ or 
$\bar{Q}^{(2)}$ and a $(2,2)$ tensor under both algebras.
The $L$-states of the $g_1$ and $g_2$ bands exhaust, respectively, 
the $(2N,0)$ and $(0,2N)$ irrep of SU(3) and $\bsu3$.
Consequently, $Q^{(2)}$ or $\bar{Q}^{(2)}$ cannot contribute to $E2$
transitions between these bands since, as generators, they cannot
mix different irreps of the respective algebras.
The $(2,2)$ tensor part of $\hat{T}(E2)$ can connect the $(2N,0)$
irrep of $g_1$ only with the $(2N-4,2)$ component in $g_2$, and the
$(0,2N)$ irrep of $g_2$ with the $(2,2N-4)$ component in $g_1$,
but these components are vanishingly small.
These observations imply that interband
$(g_2\leftrightarrow g_1)$ $E2$ transitions, are extremely weak.
By similar arguments, $E0$ transitions in-between 
the $g_1$ and $g_2$ bands are extremely weak, 
since the relevant operator, 
$\hat{T}(E0)\propto\hat{n}_d$, is a combination of $(0,0)$ and $(2,2)$ 
tensors under both algebras. 
Accordingly, 
the $L=0$ bandhead state of the higher ($g_2$) band, 
cannot decay by strong $E2$ or $E0$ transitions to the lower $g_1$ band, 
hence, as depicted schematically in Fig.~\ref{fig8-isomer},
displays characteristic features of an isomeric state.

\section{Partial dynamical symmetry and proton-neutron shapes}
\label{sec:pds-ibm2}

Shell model foundations of the IBM~\cite{IacTal87} and the
observation of mixed-symmetry states in nuclei~\cite{Richter10}, 
necessitate the introduction of the interacting proton-neutron boson
model~\cite{arima77,otsuka78} (named IBM-2 to distinguish it
from the original version which retained the name IBM-1).
The building blocks of the IBM-2 are monopole and quadrupole bosons, 
$\{s^{\dag}_{\rho},\,d^{\dag}_{\rho,m}\}$, 
of proton type ($\rho=\pi$) and of neutron type $(\rho=\nu$), 
representing pairs of identical valence nucleons. 
Number conserving bilinear combinations of operators in each set 
comprise the ${\rm U}_{\rho}(6)$ algebra as in Section~\ref{sec:intro},
and bosons of different types commute. Since the separate proton- 
and neutron- boson numbers, 
$\hat{N}_{\pi}$ and $\hat{N}_{\nu}$, are conserved, 
the appropriate spectrum generating algebra of the model 
is ${\rm U}_{\pi}(6)\times {\rm U}_{\nu}(6)$.
Subalgebras can be constructed with the aid of the individual subalgebras, 
${\rm U_{\rho}(6)},\,{\rm U_{\rho}(5)},\,{\rm SU_{\rho}(3)},\,
\overline{{\rm SU_{\rho}(3)}},\,
{\rm SO_{\rho}(6)},\, {\rm SO_{\rho}(5)},\,{\rm SO_{\rho}(3)}$. 
For instance, for a given algebra $G_{\rho}$, 
with generators ${\cal G}_{\rho}$, there is a combined algebra 
$G_{\pi+\nu}$, with generators ${\cal G}_{\pi} + {\cal G}_{\nu}$. 

The dynamical symmetries of the IBM-2 are obtained by
identifying the lattices of embedded algebras starting
with ${\rm U_{\pi}(6)}\times {\rm U_{\nu}(6)}$
and ending with the symmetry algebra ${\rm SO_{\pi+\nu}(3)}$.
A typical DS chain in the IBM-2 and related basis have
the form~\cite{ibm},
\bsub
\ba
&&{\rm U_{\pi}(6)\times U_{\nu}(6)\supset G_{\pi}\times G_{\nu}
\supset G_{\pi+\nu} \supset \ldots \supset SO_{\pi+\nu}(3)}\\
&&
\ket {[N_{\pi}] , [N_{\nu}] ;
\lambda_{\pi} , \lambda_{\nu} ,
\lambda_{\pi\nu} , \ldots, L} ~.\qquad 
\ea
\label{ds-G-pinu}
\esub

The energy surface,
$E_{N_{\pi},N_{\nu}}(\beta_{\rho},\gamma_{\rho},\Omega)$,
is now a function of four shape
variables ($\beta_{\pi},\gamma_{\pi},\beta_{\nu},\gamma_{\nu}$),
and three Euler angles ($\Omega$) of the relative
orientations between the proton and neutron
shapes~\cite{LevKir90,GinLev92}.
It is obtained by the expectation value of the
Hamiltonian in an intrinsic state, 
\ba
\ket{\beta_{\pi},\gamma_{\pi},\beta_{\nu},\gamma_{\nu},
  \Omega; N_{\pi},N_{\nu}}
&=&
\ket{\beta_{\pi},\gamma_{\pi};N_{\pi}}
\ket{\beta_{\nu},\gamma_{\nu},\Omega;N_{\nu}} ~,
\label{int-state-IBM2}
\ea
which now involves a product of a proton-condensate with
deformations $(\beta_{\pi},\gamma_{\pi})$,
defined as
in Eq.~(\ref{int-state}), and a rotated (by $\Omega$)
neutron condensate with deformations
$(\beta_{\nu},\gamma_{\nu})$.
The global minimum of the energy surface,
$(\bar{\beta}_{\pi},\bar{\beta}_{\nu},\bar{\gamma}_{\pi},
\bar{\gamma}_{\nu},\bar{\Omega})$,
define the equilibrium proton-neutron shape associated with
a given IBM-2 Hamiltonian.

The construction of IBM-2 Hamiltonians with PDS, follows
the general algorithm, by considering tensor operators
which annihilate subsets of states in particular irreps
of the leading subalgebras,
$G_{\pi}\times G_{\nu} \supset G_{\pi+\nu}$, 
in the chain~(\ref{ds-G-pinu}),
\ba
\hat{T}_{\alpha}\,\ket{[N_{\pi}],[N_{\nu}];
  \lambda_\pi\!=\!\Lambda_1 , 
  \lambda_{\nu} \!=\! \Lambda_2 ,
  \lambda_{\pi\nu} \!=\! \Lambda_{12},\ldots, L} = 0 ~.
\ea
The PDS Hamiltonian then has the form as in
Eq.~(\ref{H-PDS}),
$\hat{H}_{\rm PDS} \!=\! \hat{H} \!+\! \hat{H}_c
\!=\! \hat{H}_{\rm DS} \!+\! \hat{V}_0$, 
where $\hat{H}$ and $\hat{H}_c$ are
the intrinsic and collective parts,
Eqs.~(\ref{H-normal}) and (\ref{H-col})
respectively, and $\hat{H}_{\rm DS}$ is the DS Hamiltonian
for the chain~(\ref{ds-G-pinu}).

\section{SU(3) PDS and aligned axially-deformed
  proton-neutron shapes}
\label{pds-su3pn}

The DS chain in the IBM-2, related basis and equilibrium
deformations, appropriate to the dynamics of aligned
prolate-deformed proton-neutron shapes are
\bsub
\ba
&&
{\rm U_{\pi}(6)\times U_{\nu}(6) \supset U_{\pi+\nu}(6) \supset
  SU_{\pi+\nu}(3) \supset SO_{\pi+\nu}(3)} ~,\qquad
\label{su3pn-chain}\\
&&
\ket{[N_{\pi}],[N_{\nu}]; [N_1,N_2],(\lambda,\mu),K,L}
\;\; , \;\;\ (\beta_{\pi}\!=\!\beta_{\nu}\!=\!\sqrt{2},
\bar{\gamma}_{\pi}\!=\!\bar{\gamma}_{\nu}\!=\!0,
\bar{\Omega}\!=\!0) ~.\qquad
\label{su3pn-basis}
\ea
\label{su3pn}
\esub
For a given irrep of
${\rm U}_{\pi}(6)\times {\rm U}_{\nu}(6)$, 
characterized by $N_{\pi}$ and $N_{\nu}$, 
the allowed irreps of ${\rm U}_{\pi+\nu}(6)$ are 
$[N_1,N_2] = [N_{\pi} + N_{\nu} -k, k]$, 
where $k=0,1,\ldots, {\rm min}\{N_{\pi},\, N_{\nu}\}$. 
Instead of $[N_1,N_2]$ one can use the quantities
$N\!=\!N_{\pi} + N_{\nu}$ and
$F\!=\!\tfrac{1}{2}(N_1-N_2)=\tfrac{1}{2}N - k$, 
where $(F,N)$ are the irrep labels  
of the $SU_{F}(2)\times U_{N}(1)$ algebra, which is dual to
${\rm U}_{\pi+\nu}(6)$. Here $SU_{F}(2)$ is the F-spin
algebra~\cite{arima77}, with generators
$\hat{F}_{+} \!=\! s^{\dag}_{\pi} s_{\nu}
+ d^{\dag}_{\pi}\cdot \tilde d_{\nu},\,
\hat{F}_{-} \!=\! (\hat{F}_{+})^{\dag}$,
$\hat{F}_{0} \!=\! (\hat{N}_{\pi} - \hat{N}_{\nu})/2$
and $U_{N}(1)$ is generated by
$\hat{N}\!=\!\hat{N}_{\pi} \!+\! \hat{N}_{\nu}$.
The basis states of Eq.~(\ref{su3pn}) can
equivalently be denoted by
$\ket{[N_{\pi}],[N_{\nu}];F\!=\!\tfrac{1}{2}N-k,(\lambda,\mu),K,L}$
and have $F_z \!=\! \tfrac{1}{2}(N_{\pi}-N_{\nu})$.
States with $N_2\!=\!0$ ($k\!=\!0$) have maximal F-spin,
$F_{\max}=\tfrac{1}{2}N$, and are fully symmetric
with respect to the interchange of $\pi$ and $\nu$ bosons, while
states with $N_2\!\neq\! 0$ ($k\!\neq\! 0$) have $F< F_{max}$,
are non-symmetric, and are referred to as mixed-symmetry states.

The linear Casimir of ${\rm U_{\pi+\nu}(6)}$,
$\hat{C}_{1}[\rm U_{\pi+\nu}(6)] \!=\!\hat{N}_{\pi}+\hat{N}_{\nu}$,
has eigenvalues $N$, hence no effect on excitation energies.
The quadratic Casimir and eigenvalues are given by
\bsub
\ba
\hat{C}_{2}[\rm U_{\pi+\nu}(6)] &=&
{\textstyle\tfrac{1}{2}\hat{N}(\hat{N}+8)
+ 2{\boldmath \hat{F}^2} ~,}\\
\langle\,\hat{C}_{2}[\rm U_{\pi+\nu}(6)]\,\rangle &=&
N_1(N_1+5) + N_2(N_2+3) =
\tfrac{1}{2}N(N+8) + 2F(F+1) ~.\qquad
\ea
\esub
The quadratic Casimir of ${\rm SU_{\pi+\nu}(3)}$,
$\hat{C}_{2}[\rm SU_{\pi+\nu}(3)]$, has the same form as
Eq.~(\ref{C2}), but with $Q^{(2)}\!=\! Q^{(2)}_{\pi} + Q^{(2)}_{\nu}$
and $L^{(1)} \!=\! L^{(1)}_{\pi} + L^{(1)}_{\nu}$,
and eigenvalues $f_{2}(\lambda,\mu)$, Eq.~(\ref{f2}).
Similarly, $\hat{C}_{2}[\rm SO_{\pi+\nu}(3)] \!=\!
L^{(1)}\cdot L^{(1)}$ with eigenvalues $L(L+1)$.

The DS spectrum associated with the chain~(\ref{su3pn}), consists of
rotational bands arranged in SU(3) $(\lambda,\mu)$-multiplets
with prescribed ${\rm U_{\pi+\nu}(6)}$ (F-spin) symmetry.
The lowest states in the symmetric sector
($[N,0],F\!=\!\tfrac{1}{2}N$),
involve the ground band with $(\lambda,\mu)=(2N,0)$
and symmetric $\beta_s(K\!=\!0)$ and $\gamma_s(K\!=\!2)$
bands with $(\lambda,\mu)=(2N-4,2)$. The lowest states
in the mixed-symmetry sector
($[N-1,1],F\!=\!F_{max}-~1 = \tfrac{1}{2}N-1$),
involve the
scissors $(K\!=\!1)$ band with
$(\lambda,\mu)\!=\!(2N-2,1)$  
and the antisymmetric $\beta_a(K\!=\!0)$ and
$\gamma_a(K\!=\!2)$ bands with $(\lambda,\mu)\!=\!(2N-4,2)$.
The pattern of spectra resembles that expected of rotations
and vibrations of a combined axially-deformed shape
composed of
aligned prolate-deformed $\pi$-$\nu$ shapes with a common
symmetry $\hat{z}$-axis. The energy surface of the Casimir
operators of the leading segment in the chain~(\ref{su3pn}),
${\rm U_{\pi+\nu}(6)\supset SU_{\pi+\nu}(3)}$,
has a single minimum at the equilibrium deformations
indicated in Eq.~(\ref{su3pn-basis}). For these values the
condensate $\ket{\bar{\beta}_{\rho}\!=\!\sqrt{2},
\bar{\gamma}_{\rho}\!=\!0,\bar{\Omega}\!=\!0;N_{\pi},N_{\nu}}$,
of Eq.~(\ref{int-state-IBM2}),
becomes a lowest weight state
in the irrep $(\lambda,\mu)=(2N,0)$ of ${\rm SU_{\pi+\nu}(3)}$
with maximal F-spin, $F_{max}=N/2$, and serves
as the intrinsic state for the ground band $g(K=0)$.

According to the procedure of Section~\ref{sec:pds-ibm2},
the construction of IBM-2 Hamiltonians with SU(3)-PDS
can be accomplished by means of the following operators,
\bsub
\ba
&&{\textstyle
P^{\dagger}_{\rho,0}\;\; = 
d^{\dagger}_{\rho} \cdot d^{\dagger}_{\rho} -       
2(s^{\dagger}_{\rho})^2} \;\;\;,\,\qquad\;\;
\textstyle{
P^{\dagger}_{\pi\nu,0} = 
\sqrt{2}\,(\,d^{\dagger}_{\pi}\cdot 
d^{\dagger}_{\nu} - 2s^{\dagger}_{\pi}s^{\dagger}_{\nu}\,)}
~, \qquad
\label{P0-pn}\\
&&{\textstyle
P^{\dagger}_{\rho,2m} \!=\! 
2s^{\dagger}_{\rho}d^{\dagger}_{\rho,m} 
\!+\! \sqrt{7}(d^{\dagger}_{\rho}d^{\dagger}_{\rho})^{(2)}_m} \,,\,
{\textstyle
P^{\dagger}_{\pi\nu,2m} \!=\! 
\sqrt{2} (s^{\dagger}_{\pi}d^{\dagger}_{\nu,m} 
\!+\! s^{\dagger}_{\nu}d^{\dagger}_{\pi,m})
\!+\! \sqrt{14}(d^{\dagger}_{\pi} d^{\dagger}_{\nu})^{(2)}_m},
\;\;\qquad\label{P2-pn}\\
&&\textstyle{
M^{\dag}_{2m}\;  = 
s^{\dag}_{\pi}d^{\dag}_{\nu,m}
- s^{\dag}_{\nu}d^{\dag}_{\pi,m}} \;\; , \;\;\;\quad\;\,
\textstyle{
M^{\dagger}_{Lm} = (d^{\dagger}_{\pi} d^{\dagger}_{\nu})^{(L)}_m
\;\; (L=1,3)} ~.\qquad
\label{M2-pn}
\ea
\label{P0P2M2-pn}
\esub
$P^{\dagger}_{\rho,Lm}$ ($\rho\!=\!\pi,\nu$) are the
same $L\!=\!0,2$ pairs of Eq.~(\ref{PL}).
For fixed $(Lm)$, the three operators
$P^{\dag}_{i,Lm}$ transform as $[2,0]$ under ${\rm U_{\pi+\nu}(6)}$,
{\it i.e.}, form an F-spin vector with $F\!=\!1$ and 
$(F_0\!=\!1,0,-1)\leftrightarrow (i\!=\!\pi,\pi\nu,\nu)$.
For fixed $i$, the set of six operators
($P^{\dagger}_{i,0},\,P^{\dagger}_{i,2m}$),
span the irrep $(\lambda,\mu)\!=\!(0,2)$ of ${\rm SU_{\pi+\nu}(3)}$.
The $\pi$-$\nu$ boson pairs
$M^{\dagger}_{Lm}$ $(L=1,2,3)$ transform as $[1,1]$
under ${\rm U_{\pi+\nu}(6)}$, {\it i.e.},
are F-spin scalars ($F\!=\!0$),
and span the 15-dimensional
irrep $(\lambda,\mu)\!=\!(2,1)$ of ${\rm SU_{\pi+\nu}(3)}$.
All these operators satisfy
\bsub
\ba
P_{i,L'm}\ket{[N_{\pi}],[N_{\nu}]; F\!=\!\tfrac{1}{2}N,
  (\lambda,\mu)\!=\!(2N,0),K\!=\!0,L} &=& 0 \;\; ,\,
N \!=\! (N_{\pi}\!+\!N_{\nu})/2~,\qquad
\label{P-gbandL}
\\
M_{i,L'm}\ket{[N_{\pi}],[N_{\nu}]; F\!=\!\tfrac{1}{2}N,
  (\lambda,\mu)\!=\!(2N,0),K\!=\!0,L} &=& 0 \;\; , \,
L\!=\!0,2,4,\ldots, 2N.\qquad\;\;\;\;
\label{M-gbandL}
\ea
\label{PM-gbandL}
\esub
The indicated $L$-states span the entire $(2N,0)$ irrep of
${\rm SU_{\pi+\nu}(3)}$ and form the ground band $g(K=0)$.
Equivalently,
\bsub
\ba
P_{i,L'm}\ket{g;(2N,0)K=0,F_{max},F_z; N_{\pi},N_{\nu}} &=& 0 ~, \;
F_{max} = N/2 ~,\qquad
\label{P-gband}
\\
M_{L'm}\ket{g;(2N,0)K=0,F_{max},F_z; N_{\pi},N_{\nu}} &=& 0 ~, \;
F_z = (N_{\pi}\!-\!N_{\nu})/2 ~,
\label{M-gband}
\ea
\label{PM-gband}
\esub
where,
\ba
\ket{g;(2N,0)K=0,F_{max},F_z; N_{\pi},N_{\nu}}
=\ket{\bar{\beta_{\rho}}=\sqrt{2},
\bar{\gamma}_{\rho}=0,\bar{\Omega}=0;N_{\pi},N_{\nu}} ~,
\label{gband}
\ea
is the intrinsic state~(\ref{int-state-IBM2})
for the ground band with ${\rm SU_{\pi+\nu}(3)}$
symmetry $(\lambda,\mu)=(2N,0)$. In general, the conditions
$\bar{\beta}_{\pi}\!=\!\bar{\beta}_{\nu}$,
$\bar{\gamma}_{\pi}\!=\!\bar{\gamma}_{\nu}$ and $\bar{\Omega}\!=\!0$
ensure that the intrinsic state and the $L$-states projected
from it have good F-spin, $F\!=\!F_{max}$~\cite{levginkir90}.
The relations in Eqs.~(\ref{P-gbandL}) and
(\ref{P-gband}) follow from the 
fact that the irrep  $[N\!-\!2]$ of ${\rm U_{\pi+\nu}(6)}$
does not contain the ${\rm SU_{\pi+\nu}(3)}$ irreps
obtained from the product $(2,0)\times (2N,0)$.
Similarly, the relations of Eqs.~(\ref{M-gbandL}) and
(\ref{M-gband}) follow from the fact that
$M_{L'm}$ are F-spin scalars, while the $[N\!-\!2]$ irrep
of ${\rm U_{\pi+\nu}(6)}$ does not have states with
$F=F_{max}$.

An ${\rm SU_{\pi+\nu}(3)}$ tensor expansion of the 
Hamiltonian reads,
\ba
\hat{H}  &=& 
\sum_{i}\sum_{L=0,2}
A^{(i)}_{L}P^{\dag}_{i,L}\cdot\tilde{P}_{i,L}
+ \sum_{L=1,2,3}B_{L}M^{\dag}_{L}\cdot\tilde M_{L}
+C_{2}[
P^{\dag}_{\pi\nu,2}\cdot\tilde M_{2} 
+ {\rm H.c.} ] ~,\qquad
\label{HPDS-pn}
\ea
where
${\rm H.c.}$ means Hermitian conjugate.
Several ${\rm SU_{\pi+\nu}(3)}$-scalar interactions are
contained in the expression~(\ref{HPDS-pn}). Specifically,
\bsub
\ba
\hat{\theta}_{\rho,2} &=&
-\hat C_{2}[{\rm SU_{\rho}(3)}]
+ 2\hat N_{\rho} (2\hat N_{\rho} +3) =
P^{\dagger}_{\rho,0}P_{\rho,0}
+ P^{\dagger}_{\rho,2}\cdot \tilde{P}_{\rho,2}
\;\;\; (\rho=\pi,\nu) ~,\qquad\quad
\label{theta-rho}\\
\hat{\theta}_{\pi\nu,2} &=&
-\hat C_{2}[{\rm SU_{\pi+\nu}(3)}] + 2\hat N (2\hat N+3)
= \sum_{i}\sum_{L=0,2}
P^{\dag}_{i,L}\cdot\tilde{P}_{i,L}
  + 6{\cal M}_{\pi\nu} ~,\qquad
\label{theta-pn}\\
2{\hat{\cal M}}_{\pi\nu} &=&
-\hat{C}_{2}[{\rm U}_{\pi+\nu}(6)] + \hat{N}(\hat{N}+5) = 
2[\hat {N}(\hat{N}+2)/4 - {\boldmath \hat{F}^2}]
\nonumber\\
&=&
2 ( \, W^{\dag}_2\cdot \tilde{W}_2
+ 2\sum_{L=1,3}W^{\dag}_L\cdot \tilde{W}_L\,) ~,
\label{Majorana}
\ea
\label{Cas-pn}
\esub
where $\hat{N}\!=\!\hat{N}_{\pi}+\hat{N}_{\nu}$.
The operators
$\hat C_{2}[{\rm SU_{\rho}(3)}]$~(\ref{theta-rho})
are defined as in Eq.~(\ref{C2}), in terms of
$Q^{(2)}_{\rho}$ and $L^{(1)}_{\rho}$.
The operator $\hat{\theta}_{\pi\nu,2}$~(\ref{theta-pn}),
related to $\hat{C}_{2}[{\rm SU_{\pi+\nu}(3)}]$,
and the Majorana operator~(\ref{Majorana}), related to
$\hat{C}_{2}[{\rm U_{\pi+\nu}(6)}]$, are both part of the
DS Hamiltonian of the chain~(\ref{su3pn-chain}).
The latter Hamiltonian can be transcribed in the form,
\bsub
\ba
\hat{H}_{\rm DS} &=& 
A\,\hat{\theta}_{\pi\nu,2} + B\,\hat{{\cal M}}_{\pi\nu} +
C\,\hat{C}_{2}[\rm SO_{\pi+\nu}(3)] ~.
\label{HDS-pn}\\
E_{\rm DS} &=& {\textstyle
A\, [-f_{2}(\lambda,\mu) +2N(2N+3)] +
B\,[N(N+2)/4 - F(F+1)]}
\nonumber\\
  && +\, {\textstyle C\,L(L+1) } ~.
\ea
\label{EDS-pn}
\label{HEDS-pn}
\esub
In general, the Hamiltonian of Eq.~(\ref{HPDS-pn}) is not
invariant under ${\rm SU_{\pi+\nu}(3)}$, however,
relations~(\ref{PM-gbandL})
ensure that it has a solvable subset of states,
with good symmetry, {\it i.e.},
it has ${\rm SU_{\pi+\nu}(3)}$ PDS.
It has also partial F-spin symmetry~\cite{levgin00}, since
the solvable states have good F-spin quantum number, $F\!=\!F_{max}$,
but the Hamiltonian is an F-spin scalar only 
when $A^{(\pi)}_{L}\!=\!A^{(\nu)}_{L}\!=\!A^{(\pi\nu)}_{L}\, (L\!=\!0,2)$ 
and $C_{2}\!=\!0$.

The operators $P^{\dag}_{i,L'm}$, Eqs.~(\ref{P0-pn})-(\ref{P2-pn}),
satisfy also
\bsub
\ba
&&P_{i,L'm}\ket{[N_{\pi}],[N_{\nu}]; F \!=\! \tfrac{1}{2}N\!-\!1,
  (\lambda,\mu)\!=\!(2N\!-\!2,1),K\!=\!1,L} = 0 ~,
\label{P-scL}\\
&&P_{i,L'm}
\ket{sc; (2N-2,1)K=1,F_{max}\!-\!1,F_z;N_{\pi},N_{\nu}}
= 0 ~.\qquad
\label{P-scband}
\ea
\label{P-sc}
\esub
In Eq.~(\ref{P-scL}), the states
$L\!=\!1,2,\ldots,2N\!-\!1$ span the entire $(2N-2,1)$ irrep of
${\rm SU_{\pi+\nu}(3)}$ and form the scissors band $sc(K=1)$, 
whose intrinsic state,
\bsub
\ba
&& \ket{sc; (2N-2,1)K=1,F_{max}\!-\!1,F_z;N_{\pi},N_{\nu}}
\nonumber\\
&&\qquad\qquad\qquad\qquad\qquad\;\;\;
= \tfrac{1}{\sqrt{N}}\,\Gamma^{\dagger}_{sc}\,
\ket{g;F_{max}\!-\!1,F_z; N_{\pi}\!-\!1,N_{\nu}\!-\!1} ~,
\qquad\;\,\\
&&
\Gamma^{\dagger}_{sc} =
b^{\dag}_{c,\pi}d^{\dag}_{\nu,1} - 
d^{\dag}_{\pi,1}b^{\dag}_{c,\nu} \;\;\; , \;\;\;
b^{\dag}_{c,\rho} =
\tfrac{1}{\sqrt{3}}(\sqrt{2}\,d^{\dag}_{\rho,0} + s^{\dag}_{\rho} ) ~,
\ea
\label{sc-band}
\esub
is a lowest weight state of this irrep with F\!=\!$F_{max} - 1$.
Here $\ket{g;F_{max}\!-\!1,F_z; N_{\pi}\!-\!1,N_{\nu}\!-\!1}$
is obtained from Eq.~(\ref{gband}) and
$\Gamma^{\dagger}_{sc}$ is an F-spin scalar.
Eqs.~(\ref{P-sc}) follow from the 
fact that the $[N-2,0]$ and $[N-3,1]$ irreps of ${\rm U_{\pi+\nu}(6)}$
do not contain the ${\rm SU_{\pi+\nu}(3)}$ irreps
obtained from the product $(2,0)\times (2N-2,1)$.
These relations ensure that the states of the ground and
scissors bands are solvable eigenstates of the
Hamiltonian,
\ba
\hat{H}_{\rm PDS} &=& \hat{H}_{\rm DS}
+ \sum_{i}\sum_{L=0,2}
A^{(i)}_{L}P^{\dag}_{i,L}\cdot\tilde{P}_{i,L} 
+ B\hat{{\cal M}}_{\pi\nu}
+C\,\hat{C}_{2}[\rm SO_{\pi+\nu}(3)] ~,
\label{hprime}
\ea
where $\hat{H}_{\rm DS}$ is given in Eq.~(\ref{HEDS-pn}).
Furthermore, the operators $P^{\dag}_{i,L=0}$ satisfy
\bsub
\ba
&&P_{i,0}\ket{[N_{\pi}],[N_{\nu}]; F \!=\! \tfrac{1}{2}N,
  (\lambda,\mu)\!=\!(2N\!-\!4,2),K\!=\!2,L} = 0 ~,
\label{P0-gamma-s}\\
&&P_{i,0}\ket{[N_{\pi}],[N_{\nu}]; F \!=\! \tfrac{1}{2}N\!-\!1,
  (\lambda,\mu)\!=\!(2N\!-\!4,2),K\!=\!2,L} = 0 ~.
\label{P0-gamma-a}
\ea
\label{P0-gamma-sa}
\esub
In Eqs.~(\ref{P0-gamma-s}) and ~(\ref{P0-gamma-a}),
the states have $L=2,3,\ldots,2N - 2$ and span part
of the [$(2N-4,2),F\!=\!F_{max}$]
and [$(2N-4,2),F\!=\!F_{max}\!-\!1$] irreps of
${\rm SU_{\pi+\nu}(3)}$. They form the symmetric
$\gamma_s(K\!=\!2)$ and antisymmetric
$\gamma_a(K\!=\!2)$ gamma bands, respectively.
These properties follow from the fact that
$P_{i,0}$ annihilate the intrinsic states of these bands~\cite{Pittel86},
\bsub
\ba
&&\ket{\gamma_s;(2N-4,2)K=2,F_{max}=\tfrac{1}{2}N,F_z;N_{\pi},N_{\nu}}
\nonumber\\
&&
\qquad
=\sum_{q} C^{F_{max},F_z}_{1,q;F_{max},F_z-q}\, B^{\dag}_{\gamma 2;1q}
\ket{g;F_{max}\!-\!1,F_z\!-\!q;N_{\pi}\!-\!1\!-\!q,N_{\nu}\!-\!1\!+\!q} ~,
\qquad
\label{gamma-s}\\
&&\ket{\gamma_a;(2N-4,2)K=2,
  F_{max}-1=\tfrac{1}{2}N-1,F_z;N_{\pi},N_{\nu}}
\nonumber\\
&&
\qquad
=\sum_{q} C^{F_{max}-1,F_z}_{1,q;F_{max},F_z-q}\,
B^{\dag}_{\gamma 2;1q}
\ket{g;F_{max}\!-\!1,F_z\!-\!q;N_{\pi}\!-\!1\!-\!q,N_{\nu}\!-\!1\!+\!q} ~,
\label{gamma-a}
\ea
\esub
where
$B^{\dag}_{\gamma 2;F=1,q} = \tfrac{1}{3\sqrt{2}}P^{\dag}_{i;2,2}$
$(q\!=\!1,0,-1 \!\leftrightarrow\! i\!=\!\pi,\pi\nu,\nu)$.
Here $C^{F,F_z}_{1,q;F,F_z-q}$ is a Clebsch Gordan coefficient
and $F_z= \tfrac{1}{2}(N_{\pi}-N_{\nu})$.
Altogether, Eqs.~(\ref{PM-gbandL}), (\ref{P-sc})
and (\ref{P0-gamma-sa}) ensure that the following
Hamiltonian,
\ba
\hat{H}_{\rm PDS} &=& \hat{H}_{\rm DS} +
\sum_{i}
A^{(i)}_{0}P^{\dag}_{i,0}P_{i,0} 
+ B\hat{{\cal M}}_{\pi\nu}
+C\,\hat{C}_{2}[\rm SO_{\pi+\nu}(3)] ~,
\label{H-PDS-2}
\ea
has a number of solvable bands including
the ground ($g$), scissors ($sc$), symmetric-gamma
($\gamma_s$) and anti-symmetric gamma ($\gamma_a$) bands,
\bsub
\ba
g:&&
\ket{[N_{\pi}],[N_{\nu}]; F\!=\!\tfrac{1}{2}N,
  (\lambda,\mu)\!=\!(2N,0),K\!=\!0,L}
\nonumber\\
&& \; E_{g}(L) \!=\! CL(L+1) ~,\\
sc: &&
\ket{[N_{\pi}],[N_{\nu}]; F \!=\! \tfrac{1}{2}N\!-\!1,
  (\lambda,\mu)\!=\!(2N\!-\!2,1),K\!=\!1,L}\qquad\qquad
\qquad\qquad\qquad
\nonumber\\
&& \;  E_{sc}(L) \!=\! (6A+B)N + CL(L+1) ~,\\
\gamma_s:&&
\ket{[N_{\pi}],[N_{\nu}]; F \!=\! \tfrac{1}{2}N,
  (\lambda,\mu)\!=\!(2N\!-\!4,2),K\!=\!2,L} \qquad
\nonumber\\
&& \; E_{\gamma_s}(L) \!=\! 6A(2N-1) + CL(L+1) ~,\\
\gamma_a:&&
\ket{[N_{\pi}],[N_{\nu}]; F \!=\! \tfrac{1}{2}N\!-\!1,
(\lambda,\mu)\!=\!(2N\!-\!4,2),K\!=\!2,L}
\nonumber\\
\qquad\qquad
&& \; E_{\gamma_a}(L) \!=\! 6A(2N-1) +BN + CL(L+1) ~.\qquad
\ea
\esub
Explicit expressions for electromagnetic rates of $E2$ and $M1$
transitions among these solvable states,
are available~\cite{Isacker86}.

\section{Concluding remarks}
\label{sec:concl}

We have considered several classes of two- and three-body
IBM Hamiltonians with SU(3) PDS, for single-, multiple-
and proton-neutron shapes.
In each class, subsets of solvable states with good
symmetry allow for analytic expressions of energies and
E2 rates, as well as, symmetry-based selection rules.
The cases considered include:
(a)~IBM-1 Hamiltonians with SU(3)-PDS, describing the
dynamics of a single prolate-deformed shape, with solvable
ground $g(K\!=\!0)$ and $\gamma^{k}(K\!=\!2k)$ bands.
(b)~IBM-1 Hamiltonians with SU(3)-PDS, describing the
dynamics of a single prolate-deformed shape, with solvable
ground $g(K\!=\!0)$ and $\beta(K\!=\!0)$ bands.
(c)~IBM-1 Hamiltonians with simultaneous SU(3) and $\bsu3$ PDSs,
describing the dynamics of coexisting prolate and oblate shapes,
with solvable ground bands.
(d)~IBM-2 Hamiltonians with ${\rm SU_{\pi+\nu}(3)}$ PDS,
describing the dynamics of aligned prolate-deformed $\pi$-$\nu$ shapes,
with solvable symmetric states (maximal F-spin)
comprising the ground and symmetric-gamma bands,
and solvable mixed-symmetry states (non-maximal F-spin),
comprising the scissors and anti-symmetric gamma bands.
\begin{table*}
\centering
  \caption{\label{t_number}
\small
Number of interactions in the IBM-1 and IBM-2
for the general Hamiltonians, the SU(3) DS~(\ref{su3-chain})
and ${\rm SU_{\pi+\nu}(3)}$ DS~(\ref{su3pn}) limits, and
their PDS extensions.
On the left of $\mapsto$
is the number of interactions of a given order;
this reduces to the number on the right of $\mapsto$
if one is only interested in excitation energies
in a single nucleus.}
\begin{tabular}{cccc}
\hline\noalign{\smallskip}
Order&\multicolumn{3}{c}{Number of interactions IBM-1}\\[1mm]
\cline{2-4}\noalign{\smallskip}
& General & SU(3) DS & SU(3) PDS \\ [1mm]
\hline
$1$ & $2\mapsto1$ & $1\mapsto0$ & $1\mapsto0$\\
$2$ & $7\mapsto5$ & $3\mapsto2$ & $4\mapsto3$\\
$3$ & $17\mapsto10$ & $4\mapsto1$ &
$10\mapsto6\phantom{0}$\\[1mm]
\hline
$1+2+3$ & $26\mapsto16$ & $8\mapsto3$ &
$15\mapsto9\phantom{0}$\\[1mm]
\hline\noalign{\smallskip}
Order&\multicolumn{3}{c}{Number of interactions IBM-2}\\[1mm]
\cline{2-4}\noalign{\smallskip}
& General & ${\rm SU_{\pi+\nu}(3)}$ DS &
${\rm SU_{\pi+\nu}(3)}$ PDS \\[1mm]
\hline
$1$ & $4\mapsto2$ & $2\mapsto0$ & $2\mapsto0$\\
$2$ & $25\mapsto18$ & $6\mapsto3$ & $14\mapsto11$\\[1mm]
\hline
$1+2$ & $29\mapsto20$ & $8\mapsto3$ &
$16\mapsto11$\\
\noalign{\smallskip}\hline
\end{tabular}
\end{table*} 

The analysis serves to highlight the merits
gained by using the notion of PDS.
(i)~PDS allows more flexibility by relaxing the
constraints of an exact DS and, therefore, can be applied
to more realistic situations.
(ii)~PDS supports a subset of solvable states with
good symmetry, for which related observables can be
calculated analytically and often lead to parameter-free
predictions that can be tested by comparing with the
experimental data. This was demonstrated in Fig.~\ref{fig2-Er168be2},
for $\gamma\rightarrow g$ B(E2) ratios in $^{168}$Er.
(iii)~PDS picks particular symmetry-breaking terms which
on one hand, do not destroy results previously obtained
with a DS for a segment of the spectrum, but on the other
hand, allow deviations from undesired DS predictions for
the non-solvable mixed states. This was exemplified by
the improved PDS description of the staggering in the
$\gamma$-band of $^{156}$Gd, shown in Fig.~\ref{fig4-Gd156gam-stag},
without harming the good DS description of $E2$ transitions
involving states of the ground and beta bands, shown in
Table~\ref{tab-be2}.
(iv)~PDS can be generalized to encompass several
incompatible PDSs, hence provide
symmetry-based benchmarks for the study of
shape-coexistence in nuclei. This was illustrated for
multiple prolate-oblate shapes in Section~\ref{su3-po}.
(v)~PDS provides a selection criterion for many-body
interactions. This is particularly important in the presence
of higher-order terms ({\it e.g.}, three-body) or more
degrees of freedom ({\it e.g.}, protons and neutrons),
for which the number of possible interactions in the
effective Hamiltonian grows rapidly and a
selection criterion is called for.
  
To illustrate the increase in flexibility of a Hamiltonian with
SU(3) PDS, we list in Table~\ref{t_number}
the number of interactions under the different scenarios.
Up to third order, a general rotationally invariant IBM-1
Hamiltonian has 26 independent interactions,
decreasing to 16
if one is only interested in excitation energies in
a single nucleus. (This excludes terms involving $\hat{N}$).
An IBM-1 Hamiltonian with SU(3) DS has, up to third order,
8 independent terms but 5 of them
($\hat N$, $\hat N^2$, $\hat N^3$, $\hat N\hat L^2$, and 
$\hat N\hat C_2[{\rm SU}(3)]$)
are constant in a single nucleus
or can be absorbed in an interaction of lower order,
leaving only the 3 genuinely independent terms shown in
Eq.~(\ref{H-DS}). The corresponding numbers for
an IBM-1 Hamiltonian with SU(3) PDS are 15 and 9.
The latter number agrees with the 9
terms in the Hamiltonian~(\ref{HPDShi}).
In the IBM-2 with one- and two-body terms,
the general Hamiltonian has 29 independent interactions,
decreasing to 20 relevant for excitation energies in
a single nucleus. (This excludes terms involving
$\hat{N}_{\pi}$ and $\hat{N}_{\nu}$).
An IBM-2 Hamiltonian with the ${\rm SU_{\pi+\nu}(3)}$ DS
of Eq.~(\ref{su3pn}),
has 8 independent terms decreasing to 3
for excitation energies,
upon excluding $\hat{N}_{\rho}$, $\hat{N}_{\rho}^2$
and $\hat{N}_{\pi}N_{\nu}$.
The corresponding numbers for an IBM-2 Hamiltonian
with ${\rm SU_{\pi+\nu}(3)}$ PDS are 16 and 11.
The latter number agrees with the 10 parameters in
Eq.~(\ref{HPDS-pn}),
plus the ${\rm C_{2}[SO_{\pi+\nu}(3)]}$
term of $\hat{H}_{\rm DS}$~(\ref{HEDS-pn}).
We conclude from Table~\ref{t_number}
that more than half of all possible interactions in the IBM-1
and in the IBM-2, have in fact an SU(3) PDS.

In the present contribution, we have focused on bosonic
SU(3)-type of PDSs. However, the PDS notion is not restricted
to a specific algebra nor type of constituent particles
(bosons and fermions). Examples of
PDSs associated with other DS chains of the IBM are
known~\cite{RamLevVan09,Leviatan07,LevGav17,levgin00,LevIsa02,kremer14,Lev18,Talmi97},
as well as in the interacting boson-fermion model of odd-mass
nuclei~\cite{isajollev15}
and in purely fermionic models of
nuclei~\cite{Escher00,Escher02,rowe01,rowe03,isahein08,Talmi10,isahein14}.
In addition to applications to nuclear spectroscopy, Hamiltonians
with PDS are also relevant to studies of mixed systems
with coexisting regularity and chaos~\cite{WAL93,LW96,maclev14}.

\section*{Acknowledgements}
Segments of the reported results were obtained in
collaboration with J.~E.~Garc\'\i a-Ramos (Huelva),
P.~Van~Isacker~(GANIL)
and N.~Gavrielov (HU).
This work is supported by the Israel Science Foundation (Grant 586/16).

\end{document}